\documentclass[twocolumn,superscriptaddress,longbibliography,aps,pra,preprintnumbers,
floatfix,10pt]{revtex4-1}

\usepackage{graphicx}
\usepackage{bm}
\usepackage{color}
\usepackage{epstopdf}
\usepackage{amsmath}
\usepackage{amssymb}
\usepackage{epstopdf}
\usepackage{gensymb}

\usepackage{float}

\usepackage[urlcolor=blue,colorlinks=true,citecolor=blue,linkcolor=blue,pdfstartview={FitH},bookmarks=false]{hyperref}

\graphicspath{{fig/}{./fig/}{.}}

\begin{document}

\title{Electronic and lattice properties of non-centrosymmetric superconductors ThTSi \\ (T = Co, Ir, Ni, and Pt)}

\author{A.~Ptok}
\email[e-mail: ]{aptok@mmj.pl}
\affiliation{Institute of Nuclear Physics, Polish Academy of Sciences, \\
ul. W. E. Radzikowskiego 152, PL-31342 Krak\'{o}w, Poland}

\author{K.~Domieracki}
\affiliation{Institute of Low Temperature and Structure Research, Polish Academy of Sciences, ul. Ok\'{o}lna 2, PL-50950 Wroc\l{}aw, Poland}

\author{K.~J.~Kapcia}
\affiliation{Institute of Nuclear Physics, Polish Academy of Sciences, \\
ul. W. E. Radzikowskiego 152, PL-31342 Krak\'{o}w, Poland}

\author{J.~\L{}a\.zewski}
\affiliation{Institute of Nuclear Physics, Polish Academy of Sciences, \\
ul. W. E. Radzikowskiego 152, PL-31342 Krak\'{o}w, Poland}

\author{P.~T.~Jochym}
\affiliation{Institute of Nuclear Physics, Polish Academy of Sciences, \\
ul. W. E. Radzikowskiego 152, PL-31342 Krak\'{o}w, Poland}

\author{M.~Sternik}
\affiliation{Institute of Nuclear Physics, Polish Academy of Sciences, \\
ul. W. E. Radzikowskiego 152, PL-31342 Krak\'{o}w, Poland}

\author{P.~Piekarz}
\affiliation{Institute of Nuclear Physics, Polish Academy of Sciences, \\
ul. W. E. Radzikowskiego 152, PL-31342 Krak\'{o}w, Poland}

\author{D.~Kaczorowski}
\affiliation{Institute of Low Temperature and Structure Research, Polish Academy of Sciences, ul. Ok\'{o}lna 2, PL-50950 Wroc\l{}aw, Poland}

\begin{abstract}
The theoretical studies on the electronic and lattice properties of the series of non-centrosymmetric superconductors ThTSi, where T = Co, Ni, Ir, and Pt are presented. The electronic band structure and crystal parameters were optimized within the density functional theory. 
The spin-orbit coupling leads to the splitting of the electronic bands and Fermi surfaces, with the stronger effect observed for the compounds with the heavier atoms Ir and Pt. 
The possible mixing of the spin-singlet and spin-triplet pairing in the superconducting state is discussed. 
The phonon dispersion relations and phonon density of states were obtained using the direct method. 
The dispersion curves in ThCoSi and ThIrSi exhibit the low-energy modes along the S-N-S$_0$ line with the tendency for softening and dynamic instability. 
Additionally, we calculate and analyse the contributions of phonon modes to lattice heat capacity. 
\end{abstract}

\maketitle

\section{Introduction}

Unconventional superconductors, which exhibit an anisotropic pairing
and the presence of nodes in the superconducting gap, have been extensively studied over the last decades~\cite{sigrist.ueda.91}. 
A departure from the standard BCS theory may result from  
the absence of an inversion center in the crystal structure, which gives rise to an antisymmetric spin-orbit coupling (SOC)~\cite{bauer.sigrist.12,yip.14}.
If the SOC is sufficiently large, it leads to a mixture of spin-singlet and spin-triplet components in the superconducting state~\cite{gorkov.rashba.01,smidman.salamon.17}.
Since the discovery of the first heavy fermion compound CePt$_3$Si~\cite{bauer.hilscher.04},
a great number of non-centrosymmetric superconductors were found and investigated~\cite{akazawa.hidaka.04,togano.badica.04,klimczuk.ronning.07,bauer.khan.09,bonalde.ribeiro.2011,nishikubo.yoshihiro.11,singh.hillier.14,sun.enayata.15,barker.singh.15,sakano.okawa.15,li.xu.18,carnicom.xie.18}.

\begin{figure}[!b]
\centering
\includegraphics[width=\linewidth]{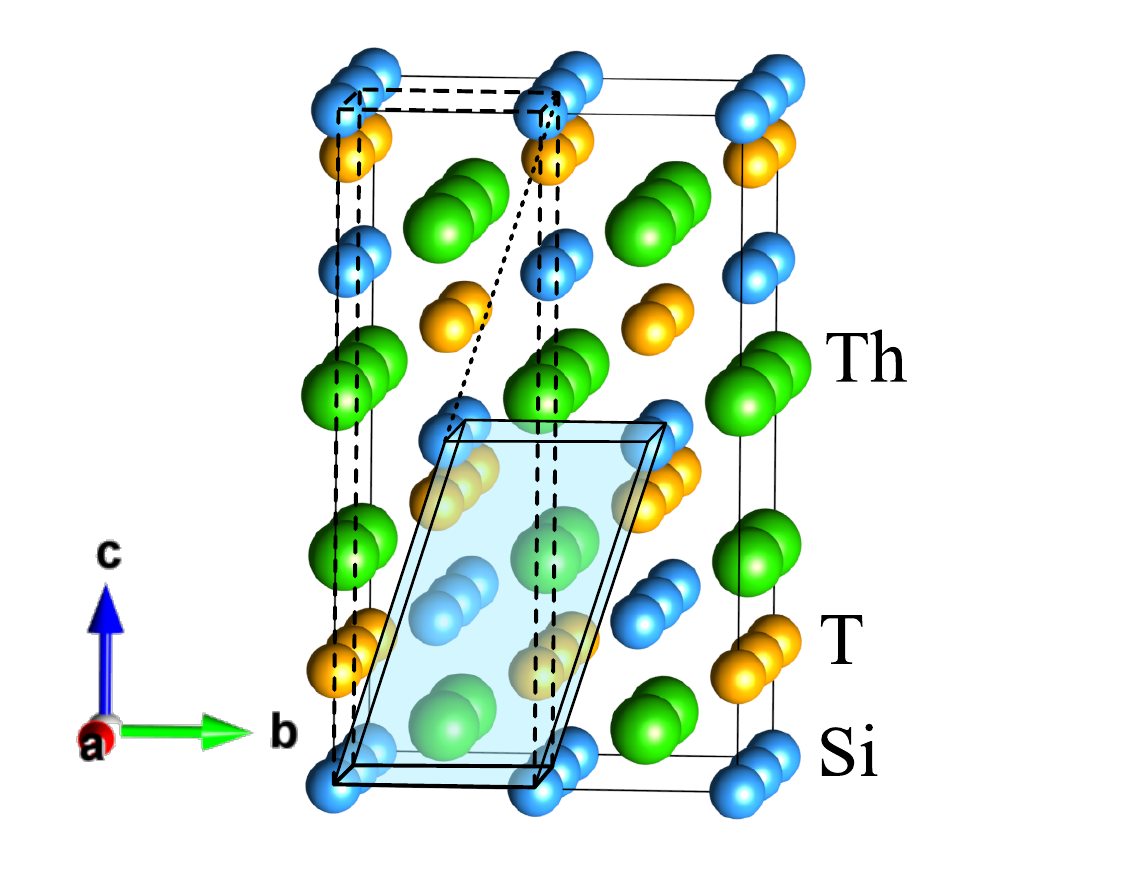} 
\caption{
\label{fig.cell}
Relation between the conventional standard cell and the primitive unit of the ThTSi compounds (marked by the bold dashed line and the solid line with the transparent filling, respectively).
The image was rendered using {\sc VESTA} software~\cite{vesta}.
}
\end{figure}

The compounds from the ThTSi family (where T = Co, Ir, Ni, and Pt) crystallize with a non-centrosymmetric tetragonal structure of the LaPtSi-type (space group {\it I4$_{1}md$}, No. 109)~\cite{klepp.parthe.82,zhong.ng.85,kneidinger.michor.13,palazzese.landaeta.18} (see Fig.~\ref{fig.cell}).
Below $T_c=2.95$~K, ThCoSi exhibits type-II moderate-coupling superconductivity with substantial Pauli pair-breaking effect~\cite{domieracki.kaczorowski.16}.
Low-temperature measurements of the zero-field specific heat shows a proportionality $C \sim T^2$, instead of an
exponential temperature dependence predicted by the BCS theory. 
This finding suggests the existence of line nodes in the superconducting gap.
The experimental characterization of ThNiSi indicates weak-coupling type-II superconductivity below $T_c=0.84$ K with somewhat
abnormal temperature dependence of the upper critical field~\cite{domieracki.kaczorowski.18}.
Contrary to ThCoSi, the superconductivity in ThNiSi
is governed by orbital pair-breaking mechanism and can be well
described within the Ginzburg-Landau theory.
These two examples clearly show a crucial role of the transition metal atom T in determining the electronic properties of the ThTSi compounds.

\begin{table*}[!t]
\caption{
\label{tab.dft}
The calculated lattice constants and atomic positions (in fractional coordinates) obtained with (w/) or without (w/o) spin-orbit coupling (SOC) compared with the experimental data. Experimental results for compound with Co and Ni are taken from Ref.~\cite{domieracki.kaczorowski.16} and~\cite{domieracki.kaczorowski.18}, respectively, whereas these with Ir and Pt are obtained within the same method.}
\begin{ruledtabular}
\begin{tabular}{lccccccccc}
system & \multicolumn{2}{c}{ThCoSi} & \multicolumn{2}{c}{ThNiSi} & \multicolumn{2}{c}{ThIrSi} & \multicolumn{2}{c}{ThPtSi} \\
& w/o SOC  & w/ SOC     & w/o SOC  & w/ SOC    & w/o SOC  & w/ SOC     & w/o SOC  & w/ SOC  \\ \hline 
$a$ (\AA)  & 4.1060  & 4.0947  & 4.1204  & 4.1153 & 4.2340  & 4.2272  & 4.2570  & 4.2515 \\ 
$c$ (\AA)  & 13.7348 & 13.6970 & 13.7830 & 13.7658 & 14.1631 & 14.1405 & 14.2399 & 14.2217 \\ 
z$_\text{Th}$ & 0.5755  & 0.5780  & 0.5794  & 0.5795 & 0.5769  & 0.5782  & 0.5797  & 0.5806 \\ 
z$_\text{T}$  & 0.1728  & 0.1707  & 0.1711  & 0.1705 & 0.1703  & 0.1704  & 0.1708  & 0.1703 \\
\hline
 & \multicolumn{8}{c}{Experimental results} \\
\hline
$a$ (\AA) & \multicolumn{2}{c}{4.081} & \multicolumn{2}{c}{4.0697} & \multicolumn{2}{c}{4.141} & \multicolumn{2}{c}{4.154} \\
$c$ (\AA) & \multicolumn{2}{c}{14.003} & \multicolumn{2}{c}{14.089} & \multicolumn{2}{c}{14.270} & \multicolumn{2}{c}{14.582} \\
z$_\text{Th}$ & \multicolumn{2}{c}{0.585} & \multicolumn{2}{c}{0.580} & \multicolumn{2}{c}{0.551} & \multicolumn{2}{c}{0.581} \\
z$_\text{T}$ & \multicolumn{2}{c}{0.1639} & \multicolumn{2}{c}{0.1662} & \multicolumn{2}{c}{0.1622} & \multicolumn{2}{c}{0.1660}
\end{tabular}
\end{ruledtabular}
\end{table*}

In this paper, we present results of our density functional theory (DFT) studies of the structural, electronic, and phonon properties on the ThTSi superconductors. 
We analysed the effect of the SOC on electronic bands and Fermi surfaces.
Phonon dispersion relations and phonon density of states were obtained by means of the direct method. 
The tendency for unstable soft-mode behavior was found for ThCoSi and ThIrSi. 
The lattice heat capacity is calculated within the harmonic approximation and compared with the experimental data.
The contribution of atomic components and the wagging modes to the heat capacity is analysed.

The paper is organized as follows. 
The details of {\it ab initio} methods are presented in Sec.~\ref{sec.det}.
The structural properties are studied in Sec.~\ref{sec.str}.
The band structure, Fermi surfaces, and density of states are presented and analysed in Sec.~\ref{sec.ele}. 
In Sec.~\ref{sec.dyn}, the phonon dispersion curves and phonon density of states are discussed. 
Lattice heat capacity is presented in Sec.~\ref{sec.cp-cv}. 
The results are summarized in Sec.~\ref{sec.sum}.

\begin{figure}[!b]
\centering
\includegraphics[width=0.75\linewidth]{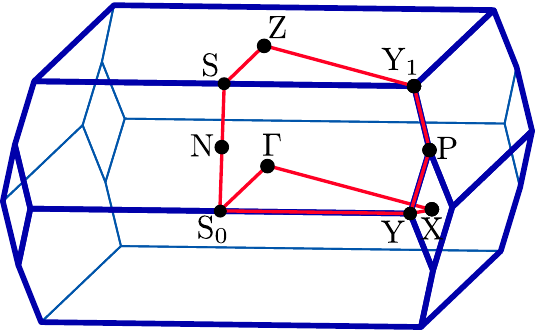}
\caption{
\label{fig.fbz}
The Brillouin zone of the $I4_{1}md$ structure with the high symmetry points~\cite{setyawan.curtarolo.10}. 
The red line demonstrates the path used for the electronic band structure presentation.
}
\end{figure}

\begin{figure*}[!t]
\centering
\includegraphics[width=\linewidth]{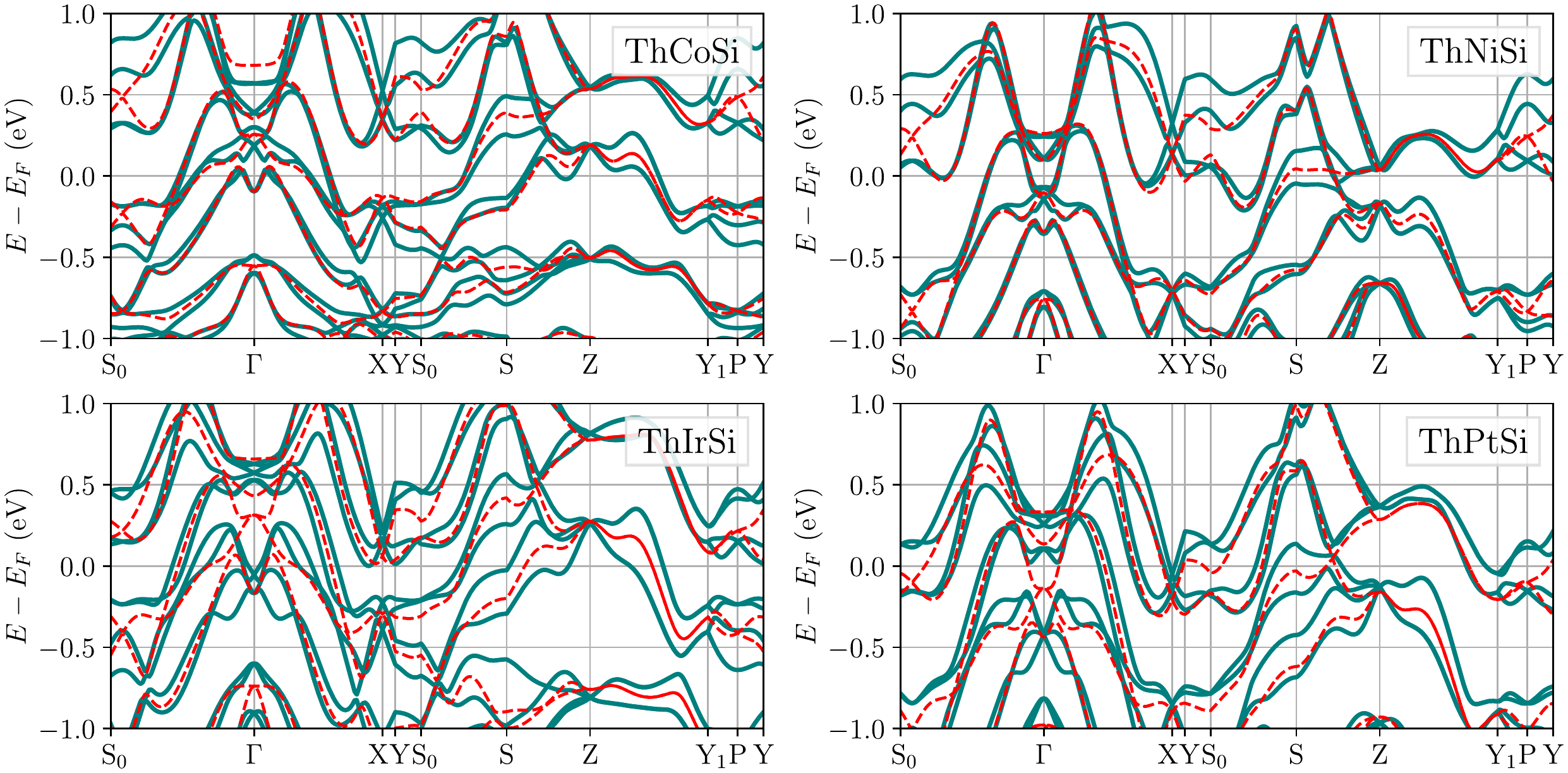}
\caption{
\label{fig.bands}
The band structure of studied systems (as labeled)
obtained in the absence and in the presence of the spin-orbit coupling (dashed red and solid green lines, respectively). 
The Fermi level is located at zero energy.
}
\end{figure*}

\section{Ab initio calculations}
\label{sec.dft}

\subsection{Calculation details}
\label{sec.det}

The first-principles calculations were performed using the projector augmented-wave (PAW) potentials~\cite{bloch.94} and generalized gradient approximation (GGA) in the Perdew, Burke and Ernzerhof (PBE) parametrization ~\cite{perdew.burke.96}, as implemented in the VASP code~\cite{kresse.hafner.94,kresse.furthmuller.96,kresse.joubert.99}.
First, the crystal structure was optimized in the conventional cell, containing two primitive unit cells (Fig.~\ref{fig.cell}). 
The calculations were performed with and without the spin-orbit coupling.
For the summation over the reciprocal space we used 15$\times$15$\times$6 {\bf k}-point grid.
The energy cut-off for plane waves expansion was set to $350$~eV.

Phonon calculations were performed in the supercell with $48$ atoms containing  2$\times$2$\times$1 conventional cells.
The phonon dispersion curves and phonon density of states (DOS) were calculated within the direct method~\cite{parlinski.li.97} implemented in the Phonon software~\cite{phonon}.
In this approach, atoms are displaced from their equilibrium positions and Hellmann-Feynman forces acting on all atoms in the supercell are calculated.
Then, the force-constant matrix elements and the dynamical matrix are constructed.
We used the displacements generated according to the thermal distribution or, 
in the case of ThIrSi, using atomic positions obtained from the molecular dynamics simulations at $T=100$~K~\cite{Hellman2013}. 
The phonon contribution to the heat capacity was calculated within the harmonic approximation.

\subsection{Crystal structures}
\label{sec.str}

All the studied ThTSi compounds were modeled by imposing the symmetry restrictions of the {\it I4$_{\it 1}md$} space group (No.~109) on the crystal structure.
The conventional crystallographic cell contains two primitive unit cells as it is shown in Fig.~\ref{fig.cell}. 
In the body-centered tetragonal structure three non-equivalent atoms: Si, T and Th are placed at the crystallographic sites: $2a(0, 0, 0)$, $2a(0, 0, z_\text{T})$ and $2a(0, 0, z_\text{Th})$, respectively.
All atoms are located in the lattice positions with the same site symmetry ($C_{2v},\, 2mm$), however, due to different mutual arrangement, thorium atoms have special coordination, other than Si and T atoms.

The lattice parameters and atomic positions obtained with and without inclusion of SOC are compared with the experimental values in Table~\ref{tab.dft}.
In all cases, the SOC inclusion results in decreasing of the lattice parameters. 
Generally, in comparison to the experimental data, our calculations overestimate $a$ and underestimate $c$ lattice constants. 
For example, for ThCoSi, $a$ is slightly longer than the experimental value $a_\text{exp}=4.081$~\AA, while the value of $c$ is smaller compared to $c_\text{exp}=14.003$~\AA~\cite{domieracki.kaczorowski.16}. 
Similar trend was found for ThNiSi, where the experimental values read $a_\text{exp}=4.070$~\AA\ and $c_\text{exp}=14.090$~\AA, respectively~\cite{domieracki.kaczorowski.18}.
Due to the heavier atoms, the lattice constants in ThIrSi and ThPtSi are significantly larger but they show similar relations between the experimental and theoretical values.
It is also worth noting that in spite of substantial differences in masses of the T atoms in the considered systems, fractional coordinates $z_\text{T}$, unconstrained by any symmetry elements, are in all these phases nearly the same.

\subsection{Electronic properties}
\label{sec.ele}

The Brillouin zone of the $I4_{1}md$ structure together with the high symmetry points and the path along which the electronic band structure was calculated are presented in Fig.~\ref{fig.fbz}. 
All calculations were performed both in the absence and in the presence of the spin-orbit interaction. 
A substantial spin-orbit splitting of the band structure is clearly visible in Fig.~\ref{fig.bands}.
The degeneracy along the high-symmetry direction $\Gamma$--Z results from the tetragonal symmetry. 
However, SOC removes the degeneracy of some energy levels at the high-symmetry points and between them. 
For example, one clearly observes this effect along the Z--Y$_{1}$ direction, where the absence of the spin-orbit interaction results in a quadruple degeneracy of electronic states (dashed red lines). 
Turning on the spin-orbit coupling reduces the degeneracy, which is twofold in this case (solid green lines).
From the observed splitting, the value of the spin-orbit coupling for a chosen momentum can be evaluated.  
These compounds are characterized by the relatively large value of the spin-orbit coupling (around $0.5$~eV) in relation to the other compounds of this class~\cite{palazzese.landaeta.18}.
The existence of such a strong spin-orbit coupling can induce effectively a mixing of the singlet and triplet superconducting pairing~\cite{gorkov.rashba.01,seo.han.12,ptok.rodrigez.18}.
However, a spin-singlet component of pairing still dominates in these compounds~\cite{kneidinger.michor.13}.
Moreover, for each compound at the point Y$_{1}$ (which connects the Z--Y$_{1}$ and Y$_{1}$--P directions), one can notice some splitting of each band.
Due to the crystal symmetry, the degeneracy along the Y$_{1}$--P line is twice lower than along the Z--Y$_{1}$ line.

\begin{figure}[!t]
\centering
\includegraphics[width=\linewidth]{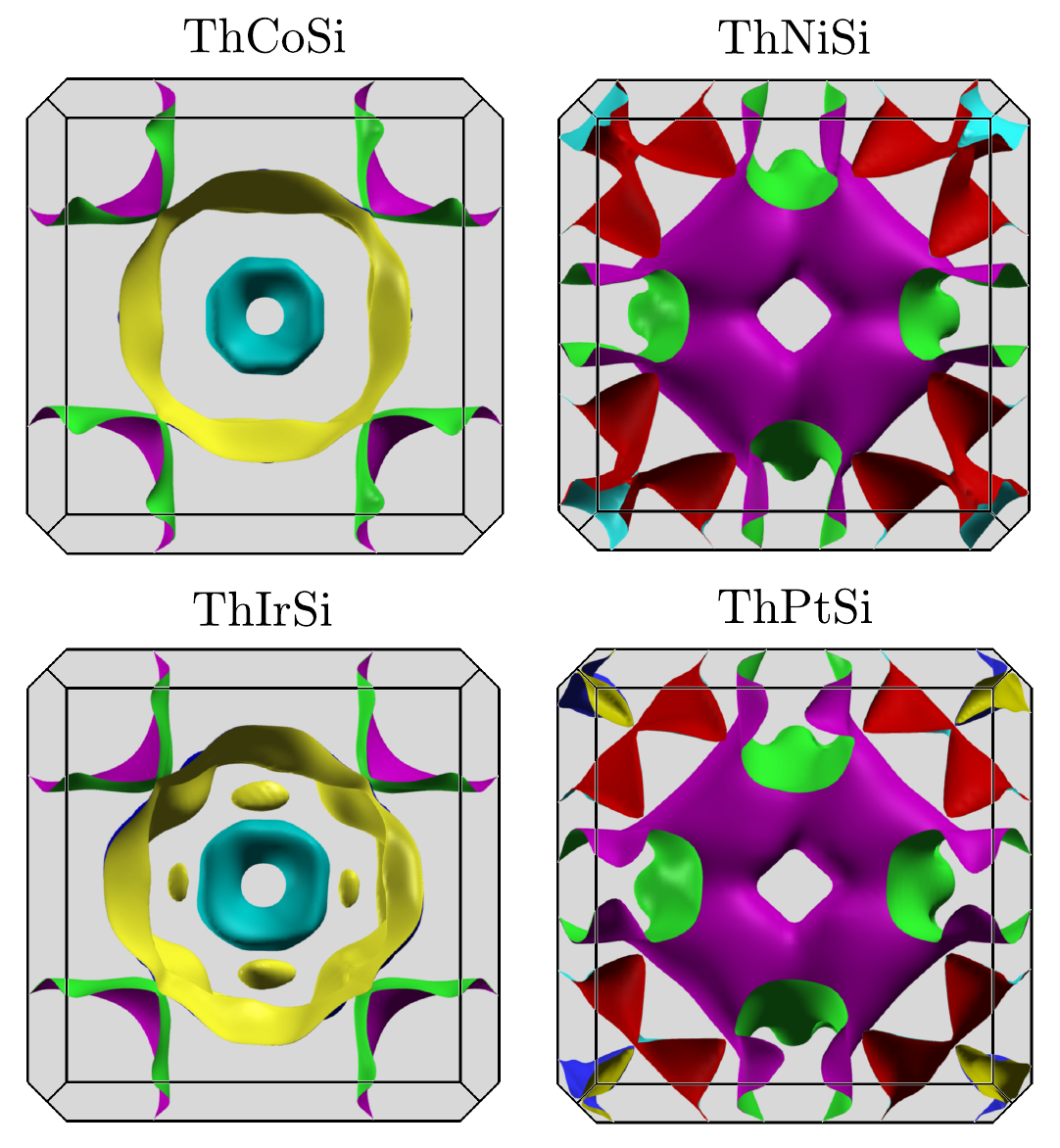}
\caption{
\label{fig.fs_nosoc}
The Fermi surfaces of studied systems (as labeled) obtained in 
the absence of the spin-orbit coupling.
}
\end{figure}

\begin{figure}[!t]
\centering
\includegraphics[width=\linewidth]{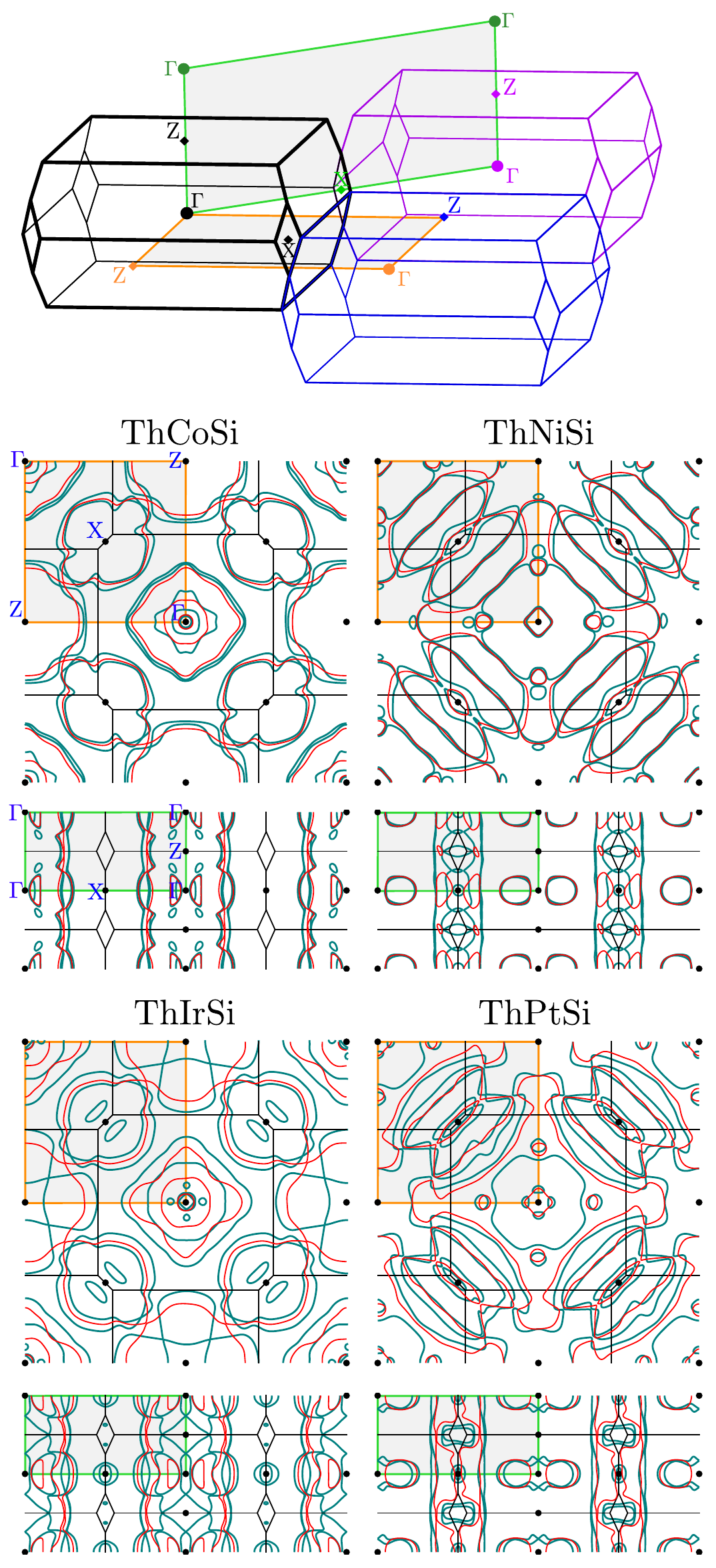}
\caption{
\label{fig.fs_cut}
Cross-sections of the Fermi surfaces of studied systems
(as labeled) in two chosen planes shown in the top panel (cf. with Fig.~\ref{fig.fbz}).
Results obtained in the absence and in the presence of the spin-orbit coupling (red and green lines, respectively).
}
\end{figure}

\begin{figure*}[!t]
\centering
\includegraphics[width=\linewidth]{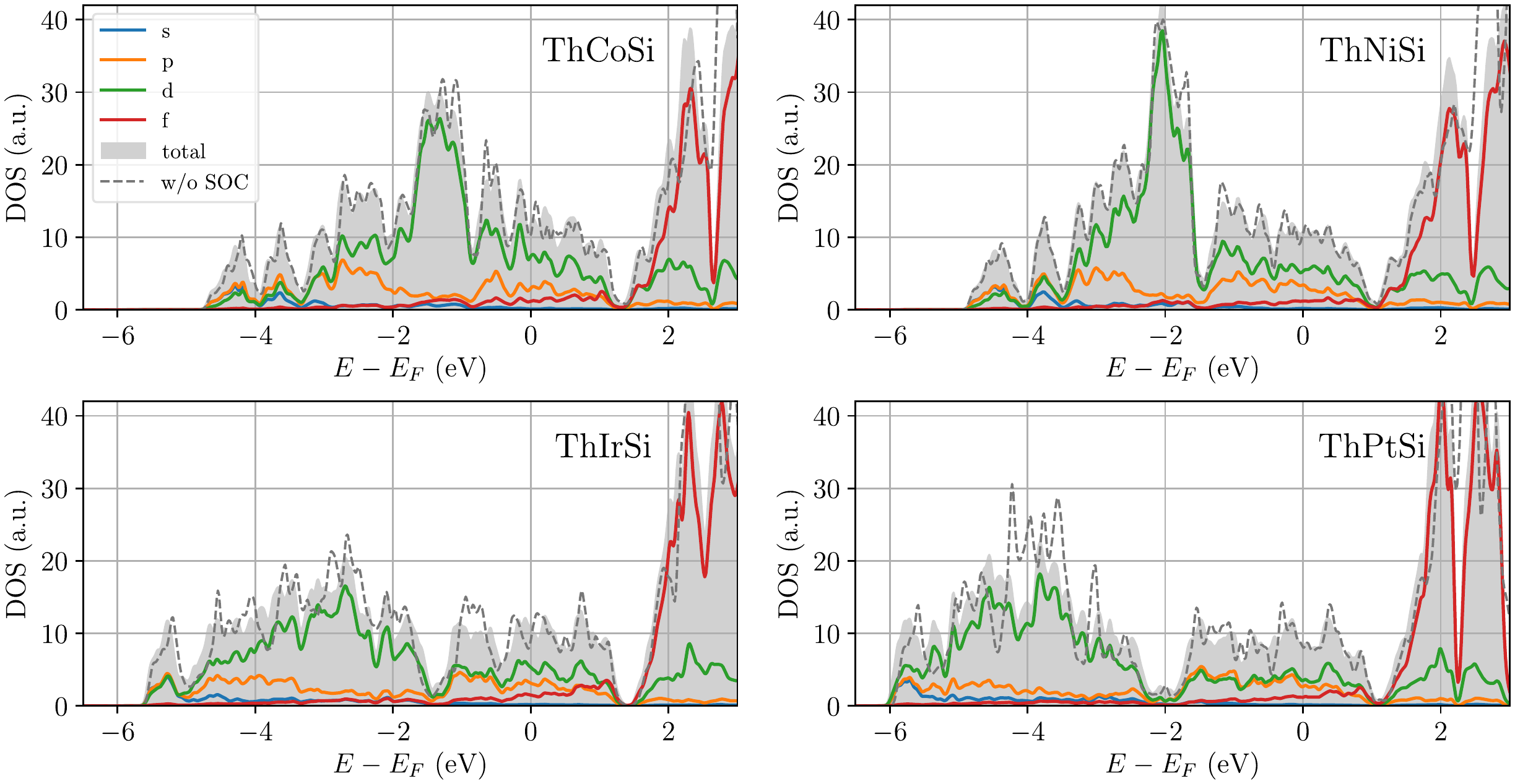}
\caption{
\label{fig.dos}
The density of states (DOS) with projections on the orbitals (as indicated in labels)
for the studied systems (as labeled). 
The Fermi level is located at zero energy.
}
\end{figure*}

The Fermi surfaces (FSs) of the studied compounds obtained in the absence of the spin-orbit coupling are shown in Fig.~\ref{fig.fs_nosoc}. 
Remarkably, FSs of ThCoSi  are very similar to those of ThIrSi, while FSs of ThNiSi are almost the same as those of ThPtSi. 
The pairs of atoms, Co and Ir as well as Ni and Pt, belong to the same groups of the periodic table of elements.
The main difference in FSs caused by an exchange of Co into Ir or Ni into Pt is associated with an emergence of new small Fermi pockets around the $\Gamma$ or X point, respectively, i.e., the occurrence of Lifshitz transition~\cite{lifshitz.60}.

An impact of the spin-orbit coupling on the Fermi surface is shown in Fig.~\ref{fig.fs_cut}.
Red and green lines correspond to the cross-sections of the Fermi surfaces by chosen planes in the case of the absence and the presence of the spin-orbit coupling, respectively.
The relatively small splitting of the profile is observed for compounds containing Co and Ni. 
However, this effect seems to be slightly larger for ThNiSi than for ThCoSi, especially, within the plane marked in Fig.~\ref{fig.fs_cut} by an orange rectangle. 
This is obviously in accordance with the results presented in Fig.~\ref{fig.bands}, which demonstrate that the magnitude of the spin-orbit splitting at the Fermi level depends strongly on a direction in the Brillouin zone. 
For systems that contain a heavy T atom, (i.e. Ir and Pt), the differences between the Fermi surfaces obtained in the absence and in the presence of the spin-orbit interaction are much more pronounced.

The total and orbital-projected electron densities of states (DOS) are presented in Fig.~\ref{fig.dos}. 
A common feature of the investigated compounds is a principal role of the $d$-type orbitals.
In the case of ThCoSi and ThNiSi, the $d$-states dominate at energies from $-2$ to $-1$~eV and from $-2.5$ to $-1.5$~eV, respectively, while in the case of ThIrSi and ThPtSi, they are mainly situated below $-2$~eV.
In all the compounds, the $5f$-orbital electron states are concentrated far above the Fermi level (approximately above $2$~eV)
The spin-orbit coupling included in the calculations only slightly changes the density of electron states
and both values of DOS at the Fermi level are almost equal.
The effect can be inferred by comparing gray dashed lines and gray areas, which correspond to total DOS in the absence and in the presence of the spin-orbit coupling, respectively.

\subsection{Lattice dynamics}
\label{sec.dyn}

\begin{figure*}[!t]
\begin{center}
\includegraphics[width=\linewidth]{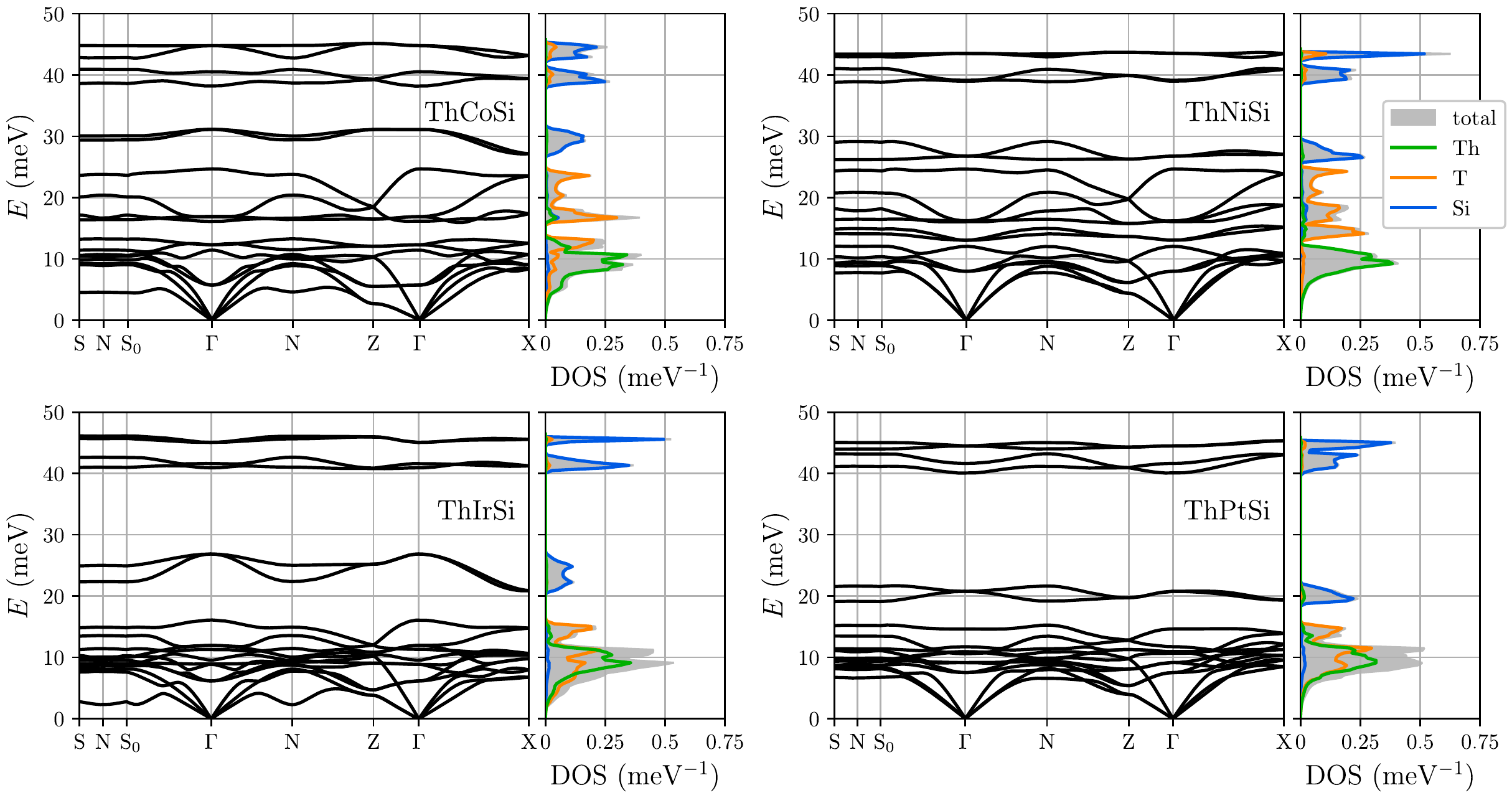}
\end{center}
\caption{
Phonon dispersion curves and density of states of studied systems (as labeled). 
}
\label{fig.phonons}
\end{figure*}

In Fig.~\ref{fig.phonons}, we compare the phonon dispersion relations and phonon DOS calculated without the spin-orbit coupling for all four compounds.
The obtained phonon frequencies are real showing dynamical stability of the crystals.
However, the lowest energy phonon branch reaches a very low energy level along the S-N-S$_0$ line in ThCoSi and ThIrSi.
In ThIrSi, these modes have a particularly strong tendency for softening.

The total and partial atom-projected phonon DOS are presented in the side panels in Fig.~\ref{fig.phonons}.
The phonon DOS of all compounds contains at least one energy gap separating upper Si-dominated band. Additionally, in Co, Ir and Pt compounds the lower Si-dominated band is also separated from the rest of the spectrum by a smaller, 2-5 meV gap. The top band in all compounds, except for ThPtSi, is further splited by another small gap of approximately 1~meV.

\begin{figure}[!b]
\begin{center}
\includegraphics[width=\linewidth]{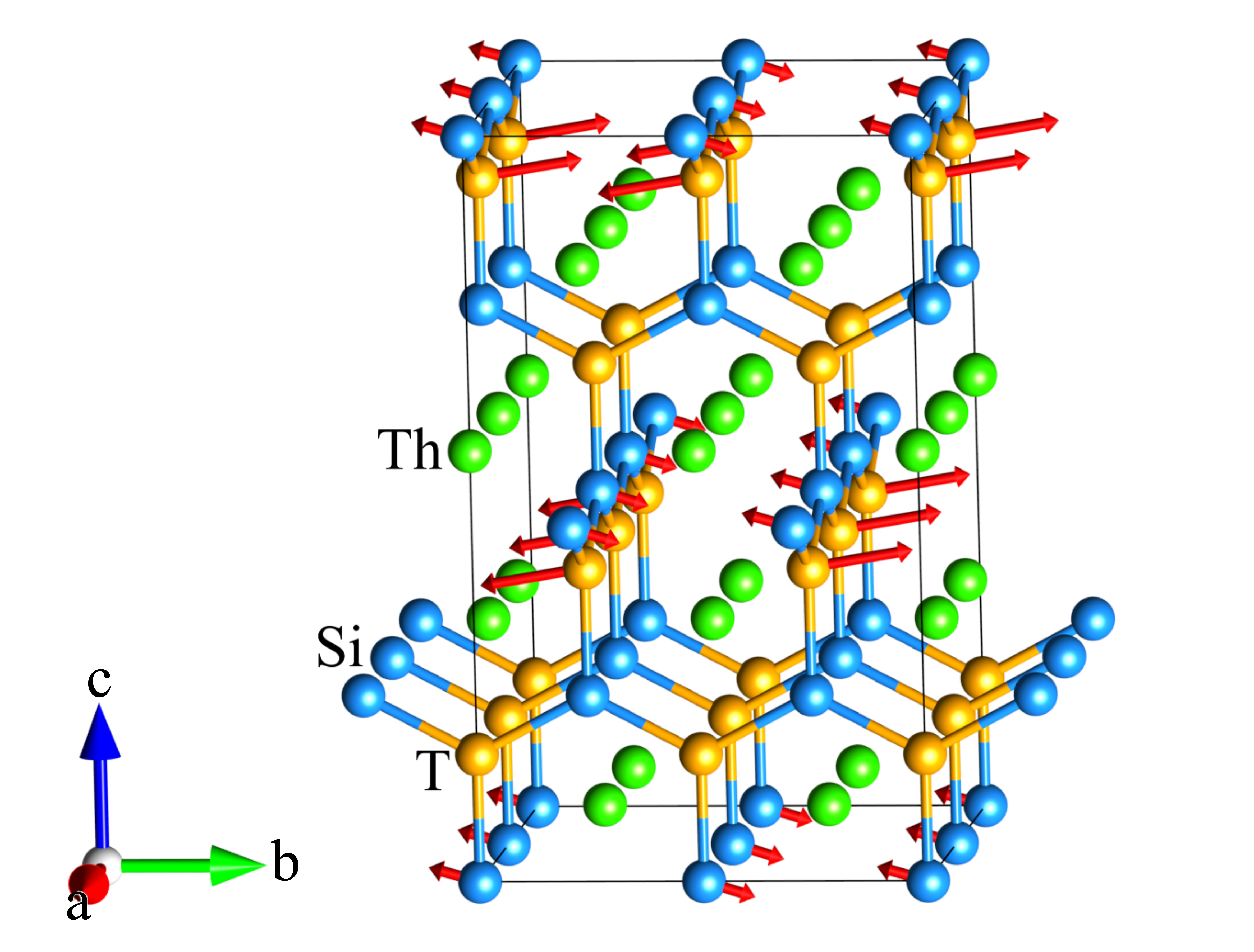}
\end{center}
\caption{An example of atomic displacements in the wagging mode with the lowest energy at the N point in ThIrSi. 
The solid lines represent schematically the three-dimensional Si-T framework based on the perpendicular T-Si-T zigzag chains.
The image was rendered using {\sc VESTA} software~\cite{vesta}.
}
\label{fig.wag}
\end{figure}

\begin{figure*}[!t]
\begin{center}
\includegraphics[width=0.75\linewidth]{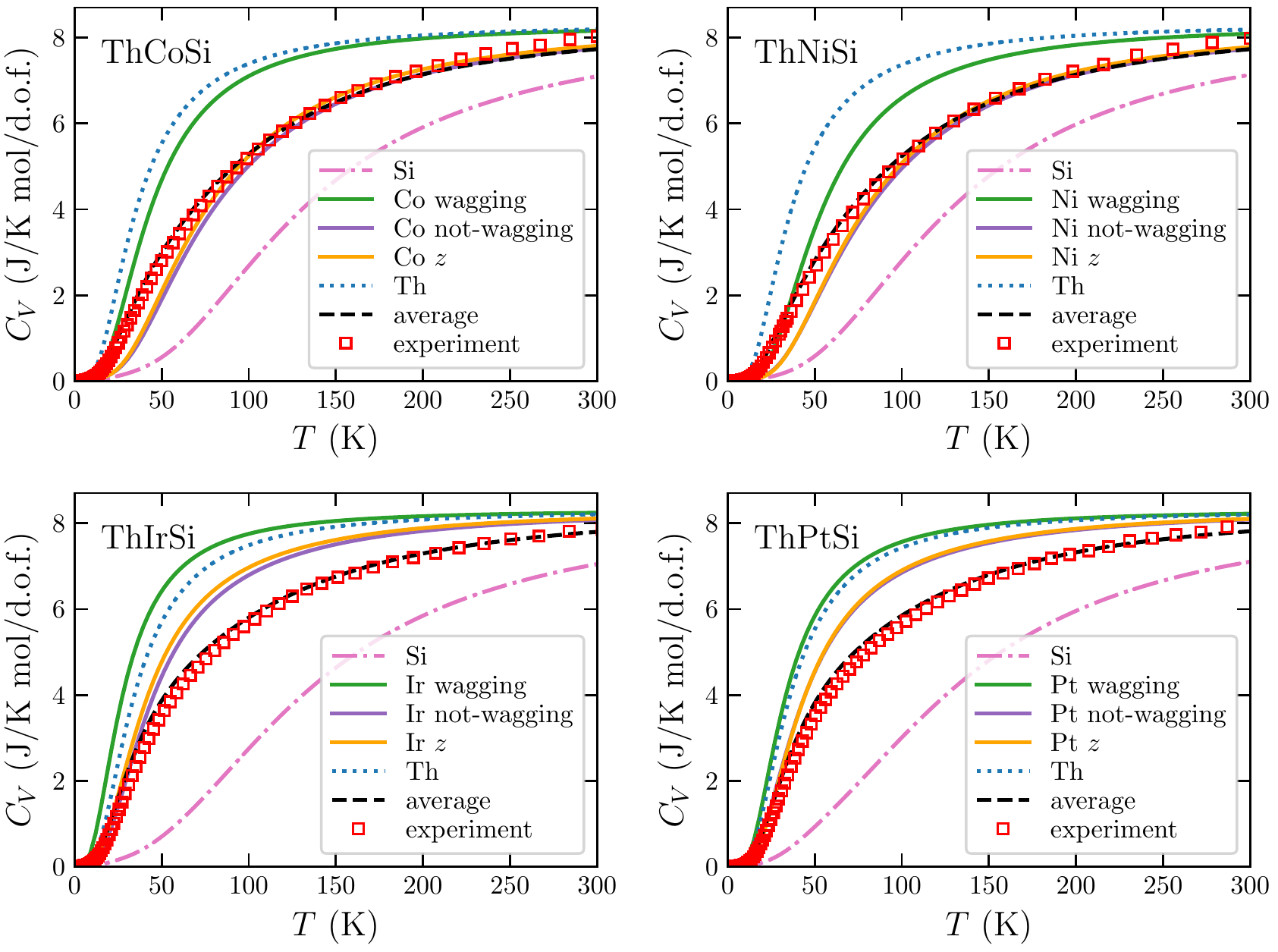}
\end{center}
\caption{The theoretical phonon heat capacity $C_V$ (black dashed lines), averaged over all degrees of freedom (d.o.f.), compared with the measured total heat capacity $C_p$ (red open squares) of studied systems. Additionally, calculated partial $C_V$ per degree of freedom are presented: averaged for Th (turquoise dotted) and Si (magenta dash-dotted) atoms as well as all components for T atoms (green, violet, and orange solid lines correspond to $xy$ wagging type, others $xy$, and $z$ contributions, respectively).
}
\label{fig.CV}
\end{figure*}

In ThCoSi, vibrations of the Th atoms dominate at energies below $13$~meV. 
The Co atoms vibrate with energies mainly between $12$--$25$~meV and have small contribution to the highest optical Si modes above $39$~meV. 
In ThNiSi, the high energy phonon branches 
are slightly shifted to lower energies in comparison with ThCoSi. 
The higher gap is increased to $\sim10$~meV, while the two lower energy gaps are reduced to about 1~meV.
The Ni atoms vibrate mainly with energies between $13$--$25$~meV, while the Si states dominate above $16$~meV.
Interestingly, in the lowest energy range (up to $12$~meV), the ThNiSi spectrum consists solely of Th vibrations, which completely separate themselves from other states. This is in  contrast to ThCoSi, where the contribution of Co is evident in this part of DOS although the masses of Co and Ni atoms are comparable. The significantly weaker binding of Th atoms in ThNiSi compound (or weaker Ni-Th interatomic forces) can explain this effect.

The situation is quite different for ThIrSi and ThPtSi where due to much larger masses of Ir and Pt, vibrations of all atoms except Si have similar frequencies and they mix together in the lower part of the phonon spectra (bottom row of Fig.~\ref{fig.phonons}). 
The Th and Pt atoms contribute mainly to the lowest energy range below $15$~meV.
There is a gap ($\sim3$~meV) which separates these states from the Si-derived bands around $20$~meV.
The higher energy gap increases to about $15$~meV in ThIrSi and $18$ meV in ThPtSi. The highest modes above $40$~meV correspond mainly to the Si atom vibrations.

The common feature of all these phonon spectra is separation of flat, dispersionless, mostly Si-derived bands located above $40$~meV. Their frequencies weakly depend on the masses of other constituent atoms. 
They are also very similar to those previously observed in the phonon spectrum of LaPtSi~\cite{kneidinger.michor.13}.

The lower vibration bands observed close to $30$~meV for T = Co or Ni and $20$~meV for T = Ir and Pt are related to the Si atoms that built the characteristic T--Si--T chains connecting Si atom with two nearest neighboring T atoms (Fig.~\ref{fig.wag}).  The T--Si--T bond angles slightly deviate from $120\degree$. Two sets of perpendicular T--Si--T zig-zag chains form the three-dimensional T--Si framework of the ThTSi crystal structure.  The interchain T-Si distances are close enough to the distances between chains. Thorium atoms are placed in the existing T--Si cages. The described mode comprise the silicon atoms vibrating out of the T--Si--T planes while the T atoms remain almost stationary. The atomic movements resemble the wagging-like modes of the Si atoms.
Similar wagging modes occur when T atoms are moving and other atoms stay almost motionless. The energies of T wagging modes, apparently lower because of larger masses of T, are the lowest energies in the partial T-derived DOS. In the case of ThCoSi and ThIrSi, they are the lowest energy modes along the S-N-S$_0$ line. A scheme of atomic displacements in the wagging mode at the N point obtained for ThIrSi is presented in Fig.~\ref{fig.wag}.

\subsection{Heat capacity}
\label{sec.cp-cv}

In Fig.~\ref{fig.CV}, the total heat capacity measured for all compounds at constant pressure ($C_p$) is compared with the phonon contribution calculated at constant volume ($C_V$).   
Experimental data were obtained by using polycrystalline samples, as it was described in Refs.~\cite{domieracki.kaczorowski.16,domieracki.kaczorowski.18}.
Below 100~K, a reasonable agreement was found for ThCoSi and ThNiSi. The largest discrepancy between the experimental and computed data is observed for ThIrSi, which is connected with the low energy modes around the S, N, and S$_0$ points.
At higher temperatures, the difference between $C_p$ and $C_V$ increases according to the formula $C_p-C_V = T \alpha^2 V/K$, where $\alpha$ and $K$ are the thermal expansion coefficient and isothermal bulk modulus, respectively.
The bifurcation between the experimental and theoretical curves, clearly observed for each compound above about 100 K, can be attributed to this effect.

The calculated individual atom contributions (determined per degree of freedom) to the total phonon heat capacity $C_V$ of the ThTSi phases are also presented in Fig.~\ref{fig.CV}.
For T atoms, three separate components: the projection on $a$ or $b$  direction due to wagging (T wagging) and non-wagging modes (T not-wagging) and $c$ direction (T~$z$) were considered.
Vibrations of the Th atoms were found very stable with well defined contribution to the heat capacity, generally independent of the T atom type. 
This finding can be directly associated with the quite symmetric coordination polyhedrons of these atoms.
On the contrary, the Si atoms yield very small contribution to $C_V$. This is a consequence
of their relatively low mass, which implies high-energetic phonon spectra, well separated from other modes. 
The situation is completely different for the T atoms, where partial $C_V$ component connected with wagging modes is visibly detached from the other ones. The contributions of the Ir and Pt atoms are even larger than those originating from the much heavier Th atoms.

\section{summary}
\label{sec.sum}

To summarize, we have presented the results of theoretical studies on the series of non-centrosymmetric superconductors ThTSi, where T = Co, Ni, Ir, and Pt. 
Using the DFT, we optimized the crystal parameters and compared them with the experimental values. For the relaxed systems, the electronic bands and Fermi surfaces
were calculated.  Similar characteristic features of the electronic structures in the compounds, which contain the T atoms from the same groups of the Periodic Table, were found.
The SOC removes the degeneracy of the electronic states and splits the spin-up and spin-down states. The stronger impact of the SOC is found for the compounds with the heavier atoms Ir and Pt.

For each ThTSi compound, we found that the Th atoms are located in quite symmetric cages built of the T and Si atoms. Their surroundings can be modified not only by changing the T atoms but also by proper doping which can tune dynamical properties of the crystal.
The calculated phonon dispersion curves and density of states indicate the dynamical stability
of all the studied materials, however, the tendency for phonon softening along the S-N-S$_0$ line in the Brillouin zone is found in ThCoSi and ThIrSi. In all investigated compounds strong separation of low-dispersion Si vibration bands have been found, which can be attributed to large mass contrast and particular arrangement of Si atoms separated and surrounded by much heavier atoms.
The lattice heat capacity was obtained and compared with the experimental data.
The analysis of the partial $C_V$ of the Th, T, and Si atoms revealed a large contribution of the wagging modes involving the T atoms, especially the heavy atoms Ir and Pt, to the heat capacity.

\begin{acknowledgments}
We thank Krzysztof Parlinski for valuable comments and discussions.
This work was supported by the National Science Centre (NCN, Poland) under grants UMO-2017/25/B/ST3/02586 (A.P., J.\L{}., P.T.J., M.S. and P.P.), and UMO-2017/24/C/ST3/00276 (K.J.K.).
\end{acknowledgments}

\bibliography{biblio}

\begin{thebibliography}{37}%
\makeatletter
\providecommand \@ifxundefined [1]{%
 \@ifx{#1\undefined}
}%
\providecommand \@ifnum [1]{%
 \ifnum #1\expandafter \@firstoftwo
 \else \expandafter \@secondoftwo
 \fi
}%
\providecommand \@ifx [1]{%
 \ifx #1\expandafter \@firstoftwo
 \else \expandafter \@secondoftwo
 \fi
}%
\providecommand \natexlab [1]{#1}%
\providecommand \enquote  [1]{``#1''}%
\providecommand \bibnamefont  [1]{#1}%
\providecommand \bibfnamefont [1]{#1}%
\providecommand \citenamefont [1]{#1}%
\providecommand \href@noop [0]{\@secondoftwo}%
\providecommand \href [0]{\begingroup \@sanitize@url \@href}%
\providecommand \@href[1]{\@@startlink{#1}\@@href}%
\providecommand \@@href[1]{\endgroup#1\@@endlink}%
\providecommand \@sanitize@url [0]{\catcode `\\12\catcode `\$12\catcode
  `\&12\catcode `\#12\catcode `\^12\catcode `\_12\catcode `\%12\relax}%
\providecommand \@@startlink[1]{}%
\providecommand \@@endlink[0]{}%
\providecommand \url  [0]{\begingroup\@sanitize@url \@url }%
\providecommand \@url [1]{\endgroup\@href {#1}{\urlprefix }}%
\providecommand \urlprefix  [0]{URL }%
\providecommand \Eprint [0]{\href }%
\providecommand \doibase [0]{http://dx.doi.org/}%
\providecommand \selectlanguage [0]{\@gobble}%
\providecommand \bibinfo  [0]{\@secondoftwo}%
\providecommand \bibfield  [0]{\@secondoftwo}%
\providecommand \translation [1]{[#1]}%
\providecommand \BibitemOpen [0]{}%
\providecommand \bibitemStop [0]{}%
\providecommand \bibitemNoStop [0]{.\EOS\space}%
\providecommand \EOS [0]{\spacefactor3000\relax}%
\providecommand \BibitemShut  [1]{\csname bibitem#1\endcsname}%
\let\auto@bib@innerbib\@empty
\bibitem [{\citenamefont {Sigrist}\ and\ \citenamefont
  {Ueda}(1991)}]{sigrist.ueda.91}%
  \BibitemOpen
  \bibfield  {author} {\bibinfo {author} {\bibfnamefont {M.}~\bibnamefont
  {Sigrist}}\ and\ \bibinfo {author} {\bibfnamefont {K.}~\bibnamefont {Ueda}},\
  }\bibfield  {title} {\enquote {\bibinfo {title} {Phenomenological theory of
  unconventional superconductivity},}\ }\href {\doibase
  10.1103/RevModPhys.63.239} {\bibfield  {journal} {\bibinfo  {journal} {Rev.
  Mod. Phys.}\ }\textbf {\bibinfo {volume} {63}},\ \bibinfo {pages} {239}
  (\bibinfo {year} {1991})}\BibitemShut {NoStop}%
\bibitem [{\citenamefont {Bauer}\ and\ \citenamefont
  {Sigrist}(2012)}]{bauer.sigrist.12}%
  \BibitemOpen
  \bibfield  {author} {\bibinfo {author} {\bibfnamefont {E.}~\bibnamefont
  {Bauer}}\ and\ \bibinfo {author} {\bibfnamefont {M.}~\bibnamefont
  {Sigrist}},\ }\href@noop {} {\emph {\bibinfo {title} {Non-centrosymmetric
  superconductors: introduction and overview}}},\ Vol.\ \bibinfo {volume}
  {847}\ (\bibinfo  {publisher} {Springer Science \& Business Media},\ \bibinfo
  {year} {2012})\BibitemShut {NoStop}%
\bibitem [{\citenamefont {Yip}(2014)}]{yip.14}%
  \BibitemOpen
  \bibfield  {author} {\bibinfo {author} {\bibfnamefont {S.}~\bibnamefont
  {Yip}},\ }\bibfield  {title} {\enquote {\bibinfo {title} {Noncentrosymmetric
  superconductors},}\ }\href {\doibase
  10.1146/annurev-conmatphys-031113-133912} {\bibfield  {journal} {\bibinfo
  {journal} {Annu. Rev. Condens. Matter Phys.}\ }\textbf {\bibinfo {volume}
  {5}},\ \bibinfo {pages} {15} (\bibinfo {year} {2014})}\BibitemShut {NoStop}%
\bibitem [{\citenamefont {Gor'kov}\ and\ \citenamefont
  {Rashba}(2001)}]{gorkov.rashba.01}%
  \BibitemOpen
  \bibfield  {author} {\bibinfo {author} {\bibfnamefont {L.~P.}\ \bibnamefont
  {Gor'kov}}\ and\ \bibinfo {author} {\bibfnamefont {E.~I.}\ \bibnamefont
  {Rashba}},\ }\bibfield  {title} {\enquote {\bibinfo {title} {Superconducting
  {2D} system with lifted spin degeneracy: Mixed singlet-triplet state},}\
  }\href {\doibase 10.1103/PhysRevLett.87.037004} {\bibfield  {journal}
  {\bibinfo  {journal} {Phys. Rev. Lett.}\ }\textbf {\bibinfo {volume} {87}},\
  \bibinfo {pages} {037004} (\bibinfo {year} {2001})}\BibitemShut {NoStop}%
\bibitem [{\citenamefont {Smidman}\ \emph {et~al.}(2017)\citenamefont
  {Smidman}, \citenamefont {Salamon}, \citenamefont {Yuan},\ and\ \citenamefont
  {Agterberg}}]{smidman.salamon.17}%
  \BibitemOpen
  \bibfield  {author} {\bibinfo {author} {\bibfnamefont {M.}~\bibnamefont
  {Smidman}}, \bibinfo {author} {\bibfnamefont {M.~B.}\ \bibnamefont
  {Salamon}}, \bibinfo {author} {\bibfnamefont {H.~Q.}\ \bibnamefont {Yuan}}, \
  and\ \bibinfo {author} {\bibfnamefont {D.~F.}\ \bibnamefont {Agterberg}},\
  }\bibfield  {title} {\enquote {\bibinfo {title} {Superconductivity and
  spin--orbit coupling in non-centrosymmetric materials: a review},}\ }\href
  {\doibase 10.1088/1361-6633/80/3/036501} {\bibfield  {journal} {\bibinfo
  {journal} {Rep. Prog. Phys.}\ }\textbf {\bibinfo {volume} {80}},\ \bibinfo
  {pages} {036501} (\bibinfo {year} {2017})}\BibitemShut {NoStop}%
\bibitem [{\citenamefont {Bauer}\ \emph {et~al.}(2004)\citenamefont {Bauer},
  \citenamefont {Hilscher}, \citenamefont {Michor}, \citenamefont {Paul},
  \citenamefont {Scheidt}, \citenamefont {Gribanov}, \citenamefont {Seropegin},
  \citenamefont {No\"el}, \citenamefont {Sigrist},\ and\ \citenamefont
  {Rogl}}]{bauer.hilscher.04}%
  \BibitemOpen
  \bibfield  {author} {\bibinfo {author} {\bibfnamefont {E.}~\bibnamefont
  {Bauer}}, \bibinfo {author} {\bibfnamefont {G.}~\bibnamefont {Hilscher}},
  \bibinfo {author} {\bibfnamefont {H.}~\bibnamefont {Michor}}, \bibinfo
  {author} {\bibfnamefont {Ch.}\ \bibnamefont {Paul}}, \bibinfo {author}
  {\bibfnamefont {E.~W.}\ \bibnamefont {Scheidt}}, \bibinfo {author}
  {\bibfnamefont {A.}~\bibnamefont {Gribanov}}, \bibinfo {author}
  {\bibfnamefont {Yu.}\ \bibnamefont {Seropegin}}, \bibinfo {author}
  {\bibfnamefont {H.}~\bibnamefont {No\"el}}, \bibinfo {author} {\bibfnamefont
  {M.}~\bibnamefont {Sigrist}}, \ and\ \bibinfo {author} {\bibfnamefont
  {P.}~\bibnamefont {Rogl}},\ }\bibfield  {title} {\enquote {\bibinfo {title}
  {Heavy fermion superconductivity and magnetic order in noncentrosymmetric
  {CePt$_{3}$Si}},}\ }\href {\doibase 10.1103/PhysRevLett.92.027003} {\bibfield
   {journal} {\bibinfo  {journal} {Phys. Rev. Lett.}\ }\textbf {\bibinfo
  {volume} {92}},\ \bibinfo {pages} {027003} (\bibinfo {year}
  {2004})}\BibitemShut {NoStop}%
\bibitem [{\citenamefont {Akazawa}\ \emph {et~al.}(2004)\citenamefont
  {Akazawa}, \citenamefont {Hidaka}, \citenamefont {Fujiwara}, \citenamefont
  {Kobayashi}, \citenamefont {Yamamoto}, \citenamefont {Haga}, \citenamefont
  {Settai},\ and\ \citenamefont {\={O}nuki}}]{akazawa.hidaka.04}%
  \BibitemOpen
  \bibfield  {author} {\bibinfo {author} {\bibfnamefont {T.}~\bibnamefont
  {Akazawa}}, \bibinfo {author} {\bibfnamefont {H.}~\bibnamefont {Hidaka}},
  \bibinfo {author} {\bibfnamefont {T.}~\bibnamefont {Fujiwara}}, \bibinfo
  {author} {\bibfnamefont {T.~C.}\ \bibnamefont {Kobayashi}}, \bibinfo {author}
  {\bibfnamefont {E.}~\bibnamefont {Yamamoto}}, \bibinfo {author}
  {\bibfnamefont {Y.}~\bibnamefont {Haga}}, \bibinfo {author} {\bibfnamefont
  {R.}~\bibnamefont {Settai}}, \ and\ \bibinfo {author} {\bibfnamefont
  {Y.}~\bibnamefont {\={O}nuki}},\ }\bibfield  {title} {\enquote {\bibinfo
  {title} {Pressure-induced superconductivity in ferromagnetic {UIr} without
  inversion symmetry},}\ }\href {\doibase 10.1088/0953-8984/16/4/l02}
  {\bibfield  {journal} {\bibinfo  {journal} {J. Phys.: Condens. Matter}\
  }\textbf {\bibinfo {volume} {16}},\ \bibinfo {pages} {L29} (\bibinfo {year}
  {2004})}\BibitemShut {NoStop}%
\bibitem [{\citenamefont {Togano}\ \emph {et~al.}(2004)\citenamefont {Togano},
  \citenamefont {Badica}, \citenamefont {Nakamori}, \citenamefont {Orimo},
  \citenamefont {Takeya},\ and\ \citenamefont {Hirata}}]{togano.badica.04}%
  \BibitemOpen
  \bibfield  {author} {\bibinfo {author} {\bibfnamefont {K.}~\bibnamefont
  {Togano}}, \bibinfo {author} {\bibfnamefont {P.}~\bibnamefont {Badica}},
  \bibinfo {author} {\bibfnamefont {Y.}~\bibnamefont {Nakamori}}, \bibinfo
  {author} {\bibfnamefont {S.}~\bibnamefont {Orimo}}, \bibinfo {author}
  {\bibfnamefont {H.}~\bibnamefont {Takeya}}, \ and\ \bibinfo {author}
  {\bibfnamefont {K.}~\bibnamefont {Hirata}},\ }\bibfield  {title} {\enquote
  {\bibinfo {title} {Superconductivity in the metal rich {Li-Pd-B} ternary
  boride},}\ }\href {\doibase 10.1103/PhysRevLett.93.247004} {\bibfield
  {journal} {\bibinfo  {journal} {Phys. Rev. Lett.}\ }\textbf {\bibinfo
  {volume} {93}},\ \bibinfo {pages} {247004} (\bibinfo {year}
  {2004})}\BibitemShut {NoStop}%
\bibitem [{\citenamefont {Klimczuk}\ \emph {et~al.}(2007)\citenamefont
  {Klimczuk}, \citenamefont {Ronning}, \citenamefont {Sidorov}, \citenamefont
  {Cava},\ and\ \citenamefont {Thompson}}]{klimczuk.ronning.07}%
  \BibitemOpen
  \bibfield  {author} {\bibinfo {author} {\bibfnamefont {T.}~\bibnamefont
  {Klimczuk}}, \bibinfo {author} {\bibfnamefont {F.}~\bibnamefont {Ronning}},
  \bibinfo {author} {\bibfnamefont {V.}~\bibnamefont {Sidorov}}, \bibinfo
  {author} {\bibfnamefont {R.~J.}\ \bibnamefont {Cava}}, \ and\ \bibinfo
  {author} {\bibfnamefont {J.~D.}\ \bibnamefont {Thompson}},\ }\bibfield
  {title} {\enquote {\bibinfo {title} {Physical properties of the
  noncentrosymmetric superconductor {Mg$_{10}$Ir$_{19}$B$_{16}$}},}\ }\href
  {\doibase 10.1103/PhysRevLett.99.257004} {\bibfield  {journal} {\bibinfo
  {journal} {Phys. Rev. Lett.}\ }\textbf {\bibinfo {volume} {99}},\ \bibinfo
  {pages} {257004} (\bibinfo {year} {2007})}\BibitemShut {NoStop}%
\bibitem [{\citenamefont {Bauer}\ \emph {et~al.}(2009)\citenamefont {Bauer},
  \citenamefont {Khan}, \citenamefont {Michor}, \citenamefont {Royanian},
  \citenamefont {Grytsiv}, \citenamefont {Melnychenko-Koblyuk}, \citenamefont
  {Rogl}, \citenamefont {Reith}, \citenamefont {Podloucky}, \citenamefont
  {Scheidt}, \citenamefont {Wolf},\ and\ \citenamefont
  {Marsman}}]{bauer.khan.09}%
  \BibitemOpen
  \bibfield  {author} {\bibinfo {author} {\bibfnamefont {E.}~\bibnamefont
  {Bauer}}, \bibinfo {author} {\bibfnamefont {R.~T.}\ \bibnamefont {Khan}},
  \bibinfo {author} {\bibfnamefont {H.}~\bibnamefont {Michor}}, \bibinfo
  {author} {\bibfnamefont {E.}~\bibnamefont {Royanian}}, \bibinfo {author}
  {\bibfnamefont {A.}~\bibnamefont {Grytsiv}}, \bibinfo {author} {\bibfnamefont
  {N.}~\bibnamefont {Melnychenko-Koblyuk}}, \bibinfo {author} {\bibfnamefont
  {P.}~\bibnamefont {Rogl}}, \bibinfo {author} {\bibfnamefont {D.}~\bibnamefont
  {Reith}}, \bibinfo {author} {\bibfnamefont {R.}~\bibnamefont {Podloucky}},
  \bibinfo {author} {\bibfnamefont {E.-W.}\ \bibnamefont {Scheidt}}, \bibinfo
  {author} {\bibfnamefont {W.}~\bibnamefont {Wolf}}, \ and\ \bibinfo {author}
  {\bibfnamefont {M.}~\bibnamefont {Marsman}},\ }\bibfield  {title} {\enquote
  {\bibinfo {title} {{BaPtSi$_{3}$}: A noncentrosymmetric {BCS}-like
  superconductor},}\ }\href {\doibase 10.1103/PhysRevB.80.064504} {\bibfield
  {journal} {\bibinfo  {journal} {Phys. Rev. B}\ }\textbf {\bibinfo {volume}
  {80}},\ \bibinfo {pages} {064504} (\bibinfo {year} {2009})}\BibitemShut
  {NoStop}%
\bibitem [{\citenamefont {Bonalde}\ \emph {et~al.}(2011)\citenamefont
  {Bonalde}, \citenamefont {Ribeiro}, \citenamefont {Syu}, \citenamefont
  {Sung},\ and\ \citenamefont {Lee}}]{bonalde.ribeiro.2011}%
  \BibitemOpen
  \bibfield  {author} {\bibinfo {author} {\bibfnamefont {I.}~\bibnamefont
  {Bonalde}}, \bibinfo {author} {\bibfnamefont {R.~L.}\ \bibnamefont
  {Ribeiro}}, \bibinfo {author} {\bibfnamefont {K.~J.}\ \bibnamefont {Syu}},
  \bibinfo {author} {\bibfnamefont {H.~H.}\ \bibnamefont {Sung}}, \ and\
  \bibinfo {author} {\bibfnamefont {W.~H.}\ \bibnamefont {Lee}},\ }\bibfield
  {title} {\enquote {\bibinfo {title} {Nodal gap structure in the
  noncentrosymmetric superconductor {LaNiC$_{2}$} from
  magnetic-penetration-depth measurements},}\ }\href {\doibase
  10.1088/1367-2630/13/12/123022} {\bibfield  {journal} {\bibinfo  {journal}
  {New J. Phys.}\ }\textbf {\bibinfo {volume} {13}},\ \bibinfo {pages} {123022}
  (\bibinfo {year} {2011})}\BibitemShut {NoStop}%
\bibitem [{\citenamefont {Nishikubo}\ \emph {et~al.}(2011)\citenamefont
  {Nishikubo}, \citenamefont {Kudo},\ and\ \citenamefont
  {Nohara}}]{nishikubo.yoshihiro.11}%
  \BibitemOpen
  \bibfield  {author} {\bibinfo {author} {\bibfnamefont {Y.}~\bibnamefont
  {Nishikubo}}, \bibinfo {author} {\bibfnamefont {K.}~\bibnamefont {Kudo}}, \
  and\ \bibinfo {author} {\bibfnamefont {M.}~\bibnamefont {Nohara}},\
  }\bibfield  {title} {\enquote {\bibinfo {title} {Superconductivity in the
  honeycomb-lattice pnictide {SrPtAs}},}\ }\href {\doibase
  10.1143/JPSJ.80.055002} {\bibfield  {journal} {\bibinfo  {journal} {J. Phys.
  Soc. Jpn.}\ }\textbf {\bibinfo {volume} {80}},\ \bibinfo {pages} {055002}
  (\bibinfo {year} {2011})}\BibitemShut {NoStop}%
\bibitem [{\citenamefont {Singh}\ \emph {et~al.}(2014)\citenamefont {Singh},
  \citenamefont {Hillier}, \citenamefont {Mazidian}, \citenamefont
  {Quintanilla}, \citenamefont {Annett}, \citenamefont {Paul}, \citenamefont
  {Balakrishnan},\ and\ \citenamefont {Lees}}]{singh.hillier.14}%
  \BibitemOpen
  \bibfield  {author} {\bibinfo {author} {\bibfnamefont {R.~P.}\ \bibnamefont
  {Singh}}, \bibinfo {author} {\bibfnamefont {A.~D.}\ \bibnamefont {Hillier}},
  \bibinfo {author} {\bibfnamefont {B.}~\bibnamefont {Mazidian}}, \bibinfo
  {author} {\bibfnamefont {J.}~\bibnamefont {Quintanilla}}, \bibinfo {author}
  {\bibfnamefont {J.~F.}\ \bibnamefont {Annett}}, \bibinfo {author}
  {\bibfnamefont {D.~McK.~Paul}\ \bibnamefont {Paul}}, \bibinfo {author}
  {\bibfnamefont {G.}~\bibnamefont {Balakrishnan}}, \ and\ \bibinfo {author}
  {\bibfnamefont {M.~R.}\ \bibnamefont {Lees}},\ }\bibfield  {title} {\enquote
  {\bibinfo {title} {Detection of time-reversal symmetry breaking in the
  noncentrosymmetric superconductor {Re$_{6}$Zr} using muon-spin
  spectroscopy},}\ }\href {\doibase 10.1103/PhysRevLett.112.107002} {\bibfield
  {journal} {\bibinfo  {journal} {Phys. Rev. Lett.}\ }\textbf {\bibinfo
  {volume} {112}},\ \bibinfo {pages} {107002} (\bibinfo {year}
  {2014})}\BibitemShut {NoStop}%
\bibitem [{\citenamefont {Sun}\ \emph {et~al.}(2015)\citenamefont {Sun},
  \citenamefont {Enayat}, \citenamefont {Maldonado}, \citenamefont {Lithgow},
  \citenamefont {Yelland}, \citenamefont {Peets}, \citenamefont {Yaresko},
  \citenamefont {Schnyder},\ and\ \citenamefont {Wahl}}]{sun.enayata.15}%
  \BibitemOpen
  \bibfield  {author} {\bibinfo {author} {\bibfnamefont {Z.}~\bibnamefont
  {Sun}}, \bibinfo {author} {\bibfnamefont {M.}~\bibnamefont {Enayat}},
  \bibinfo {author} {\bibfnamefont {A.}~\bibnamefont {Maldonado}}, \bibinfo
  {author} {\bibfnamefont {C.}~\bibnamefont {Lithgow}}, \bibinfo {author}
  {\bibfnamefont {E.}~\bibnamefont {Yelland}}, \bibinfo {author} {\bibfnamefont
  {D.~C.}\ \bibnamefont {Peets}}, \bibinfo {author} {\bibfnamefont
  {A.}~\bibnamefont {Yaresko}}, \bibinfo {author} {\bibfnamefont {A.~P.}\
  \bibnamefont {Schnyder}}, \ and\ \bibinfo {author} {\bibfnamefont
  {P.}~\bibnamefont {Wahl}},\ }\bibfield  {title} {\enquote {\bibinfo {title}
  {Dirac surface states and nature of superconductivity in noncentrosymmetric
  {BiPd}},}\ }\href {\doibase 10.1038/ncomms7633} {\bibfield  {journal}
  {\bibinfo  {journal} {Nat. Commun.}\ }\textbf {\bibinfo {volume} {6}},\
  \bibinfo {pages} {6633} (\bibinfo {year} {2015})}\BibitemShut {NoStop}%
\bibitem [{\citenamefont {Barker}\ \emph {et~al.}(2015)\citenamefont {Barker},
  \citenamefont {Singh}, \citenamefont {Thamizhavel}, \citenamefont {Hillier},
  \citenamefont {Lees}, \citenamefont {Balakrishnan}, \citenamefont {Paul},\
  and\ \citenamefont {Singh}}]{barker.singh.15}%
  \BibitemOpen
  \bibfield  {author} {\bibinfo {author} {\bibfnamefont {J.~A.~T.}\
  \bibnamefont {Barker}}, \bibinfo {author} {\bibfnamefont {D.}~\bibnamefont
  {Singh}}, \bibinfo {author} {\bibfnamefont {A.}~\bibnamefont {Thamizhavel}},
  \bibinfo {author} {\bibfnamefont {A.~D.}\ \bibnamefont {Hillier}}, \bibinfo
  {author} {\bibfnamefont {M.~R.}\ \bibnamefont {Lees}}, \bibinfo {author}
  {\bibfnamefont {G.}~\bibnamefont {Balakrishnan}}, \bibinfo {author}
  {\bibfnamefont {D.~McK.}\ \bibnamefont {Paul}}, \ and\ \bibinfo {author}
  {\bibfnamefont {R.~P.}\ \bibnamefont {Singh}},\ }\bibfield  {title} {\enquote
  {\bibinfo {title} {Unconventional superconductivity in {La$_{7}$Ir$_{3}$}
  revealed by muon spin relaxation: Introducing a new family of
  noncentrosymmetric superconductor that breaks time-reversal symmetry},}\
  }\href {\doibase 10.1103/PhysRevLett.115.267001} {\bibfield  {journal}
  {\bibinfo  {journal} {Phys. Rev. Lett.}\ }\textbf {\bibinfo {volume} {115}},\
  \bibinfo {pages} {267001} (\bibinfo {year} {2015})}\BibitemShut {NoStop}%
\bibitem [{\citenamefont {Sakano}\ \emph {et~al.}(2015)\citenamefont {Sakano},
  \citenamefont {Okawa}, \citenamefont {Kanou}, \citenamefont {Sanjo},
  \citenamefont {Okuda}, \citenamefont {Sasagawa},\ and\ \citenamefont
  {Ishizaka}}]{sakano.okawa.15}%
  \BibitemOpen
  \bibfield  {author} {\bibinfo {author} {\bibfnamefont {M.}~\bibnamefont
  {Sakano}}, \bibinfo {author} {\bibfnamefont {K.}~\bibnamefont {Okawa}},
  \bibinfo {author} {\bibfnamefont {M.}~\bibnamefont {Kanou}}, \bibinfo
  {author} {\bibfnamefont {H.}~\bibnamefont {Sanjo}}, \bibinfo {author}
  {\bibfnamefont {T.}~\bibnamefont {Okuda}}, \bibinfo {author} {\bibfnamefont
  {T.}~\bibnamefont {Sasagawa}}, \ and\ \bibinfo {author} {\bibfnamefont
  {K.}~\bibnamefont {Ishizaka}},\ }\bibfield  {title} {\enquote {\bibinfo
  {title} {Topologically protected surface states in a centrosymmetric
  superconductor {$\beta$-PdBi$_{2}$}},}\ }\href
  {http://doi.org/10.1038/ncomms9595} {\bibfield  {journal} {\bibinfo
  {journal} {Nat. Commun.}\ }\textbf {\bibinfo {volume} {6}},\ \bibinfo {pages}
  {8595} (\bibinfo {year} {2015})}\BibitemShut {NoStop}%
\bibitem [{\citenamefont {Li}\ \emph {et~al.}(2018)\citenamefont {Li},
  \citenamefont {Xu}, \citenamefont {Zhou}, \citenamefont {Jiao}, \citenamefont
  {Sankar}, \citenamefont {Zhang}, \citenamefont {Hou}, \citenamefont {Jiang},
  \citenamefont {Qian}, \citenamefont {Chen}, \citenamefont {Bangura},\ and\
  \citenamefont {Xu}}]{li.xu.18}%
  \BibitemOpen
  \bibfield  {author} {\bibinfo {author} {\bibfnamefont {B.}~\bibnamefont
  {Li}}, \bibinfo {author} {\bibfnamefont {C.~Q.}\ \bibnamefont {Xu}}, \bibinfo
  {author} {\bibfnamefont {W.}~\bibnamefont {Zhou}}, \bibinfo {author}
  {\bibfnamefont {W.~H.}\ \bibnamefont {Jiao}}, \bibinfo {author}
  {\bibfnamefont {R.}~\bibnamefont {Sankar}}, \bibinfo {author} {\bibfnamefont
  {F.~M.}\ \bibnamefont {Zhang}}, \bibinfo {author} {\bibfnamefont {H.~H.}\
  \bibnamefont {Hou}}, \bibinfo {author} {\bibfnamefont {X.~F.}\ \bibnamefont
  {Jiang}}, \bibinfo {author} {\bibfnamefont {B.}~\bibnamefont {Qian}},
  \bibinfo {author} {\bibfnamefont {B.}~\bibnamefont {Chen}}, \bibinfo {author}
  {\bibfnamefont {A.~F.}\ \bibnamefont {Bangura}}, \ and\ \bibinfo {author}
  {\bibfnamefont {X.}~\bibnamefont {Xu}},\ }\bibfield  {title} {\enquote
  {\bibinfo {title} {Evidence of s-wave superconductivity in the
  noncentrosymmetric {La$_{7}$Ir$_{3}$}},}\ }\href {\doibase
  10.1038/s41598-017-19042-x} {\bibfield  {journal} {\bibinfo  {journal} {Sci.
  Rep.}\ }\textbf {\bibinfo {volume} {8}},\ \bibinfo {pages} {651} (\bibinfo
  {year} {2018})}\BibitemShut {NoStop}%
\bibitem [{\citenamefont {Carnicom}\ \emph {et~al.}(2018)\citenamefont
  {Carnicom}, \citenamefont {Xie}, \citenamefont {Klimczuk}, \citenamefont
  {Lin}, \citenamefont {G{\'o}rnicka}, \citenamefont {Sobczak}, \citenamefont
  {Ong},\ and\ \citenamefont {Cava}}]{carnicom.xie.18}%
  \BibitemOpen
  \bibfield  {author} {\bibinfo {author} {\bibfnamefont {E.~M.}\ \bibnamefont
  {Carnicom}}, \bibinfo {author} {\bibfnamefont {W.}~\bibnamefont {Xie}},
  \bibinfo {author} {\bibfnamefont {T.}~\bibnamefont {Klimczuk}}, \bibinfo
  {author} {\bibfnamefont {J.}~\bibnamefont {Lin}}, \bibinfo {author}
  {\bibfnamefont {K.}~\bibnamefont {G{\'o}rnicka}}, \bibinfo {author}
  {\bibfnamefont {Z.}~\bibnamefont {Sobczak}}, \bibinfo {author} {\bibfnamefont
  {N.~P.}\ \bibnamefont {Ong}}, \ and\ \bibinfo {author} {\bibfnamefont
  {R.~J.}\ \bibnamefont {Cava}},\ }\bibfield  {title} {\enquote {\bibinfo
  {title} {{TaRh$_{2}$B$_{2}$} and {NbRh$_{2}$B$_{2}$}: Superconductors with a
  chiral noncentrosymmetric crystal structure},}\ }\href {\doibase
  10.1126/sciadv.aar7969} {\bibfield  {journal} {\bibinfo  {journal} {Sci.
  Adv.}\ }\textbf {\bibinfo {volume} {4}},\ \bibinfo {pages} {eaar7969}
  (\bibinfo {year} {2018})}\BibitemShut {NoStop}%
\bibitem [{\citenamefont {Klepp}\ and\ \citenamefont
  {Parth{\'{e}}}(1982)}]{klepp.parthe.82}%
  \BibitemOpen
  \bibfield  {author} {\bibinfo {author} {\bibfnamefont {K.}~\bibnamefont
  {Klepp}}\ and\ \bibinfo {author} {\bibfnamefont {E.}~\bibnamefont
  {Parth{\'{e}}}},\ }\bibfield  {title} {\enquote {\bibinfo {title} {{{{\it
  R}PtSi} phases {({\it R} = La, Ce, Pr, Nd, Sm and Gd)} with an ordered
  {ThSi$_{2}$} derivative structure}},}\ }\href {\doibase
  10.1107/S056774088200507X} {\bibfield  {journal} {\bibinfo  {journal} {Acta
  Crystallogr. B}\ }\textbf {\bibinfo {volume} {38}},\ \bibinfo {pages} {1105}
  (\bibinfo {year} {1982})}\BibitemShut {NoStop}%
\bibitem [{\citenamefont {Zhong}\ \emph {et~al.}(1985)\citenamefont {Zhong},
  \citenamefont {Ng}, \citenamefont {Chevalier}, \citenamefont {Etourneau},\
  and\ \citenamefont {Hagenmuller}}]{zhong.ng.85}%
  \BibitemOpen
  \bibfield  {author} {\bibinfo {author} {\bibfnamefont {W.~X.}\ \bibnamefont
  {Zhong}}, \bibinfo {author} {\bibfnamefont {W.~L.}\ \bibnamefont {Ng}},
  \bibinfo {author} {\bibfnamefont {B.}~\bibnamefont {Chevalier}}, \bibinfo
  {author} {\bibfnamefont {J.}~\bibnamefont {Etourneau}}, \ and\ \bibinfo
  {author} {\bibfnamefont {P.}~\bibnamefont {Hagenmuller}},\ }\bibfield
  {title} {\enquote {\bibinfo {title} {Structural and electrical properties of
  new silicides: {ThCoxSi$_{2-x}$} (0 $\leq$ x $\leq$ 1) and {ThTSi} ({T} =
  {Ni}, {Pt})},}\ }\href {\doibase 10.1016/0025-5408(85)90097-2} {\bibfield
  {journal} {\bibinfo  {journal} {Mater. Res. Bull.}\ }\textbf {\bibinfo
  {volume} {20}},\ \bibinfo {pages} {1229} (\bibinfo {year}
  {1985})}\BibitemShut {NoStop}%
\bibitem [{\citenamefont {Kneidinger}\ \emph {et~al.}(2013)\citenamefont
  {Kneidinger}, \citenamefont {Michor}, \citenamefont {Sidorenko},
  \citenamefont {Bauer}, \citenamefont {Zeiringer}, \citenamefont {Rogl},
  \citenamefont {Blaas-Schenner}, \citenamefont {Reith},\ and\ \citenamefont
  {Podloucky}}]{kneidinger.michor.13}%
  \BibitemOpen
  \bibfield  {author} {\bibinfo {author} {\bibfnamefont {F.}~\bibnamefont
  {Kneidinger}}, \bibinfo {author} {\bibfnamefont {H.}~\bibnamefont {Michor}},
  \bibinfo {author} {\bibfnamefont {A.}~\bibnamefont {Sidorenko}}, \bibinfo
  {author} {\bibfnamefont {E.}~\bibnamefont {Bauer}}, \bibinfo {author}
  {\bibfnamefont {I.}~\bibnamefont {Zeiringer}}, \bibinfo {author}
  {\bibfnamefont {P.}~\bibnamefont {Rogl}}, \bibinfo {author} {\bibfnamefont
  {C.}~\bibnamefont {Blaas-Schenner}}, \bibinfo {author} {\bibfnamefont
  {D.}~\bibnamefont {Reith}}, \ and\ \bibinfo {author} {\bibfnamefont
  {R.}~\bibnamefont {Podloucky}},\ }\bibfield  {title} {\enquote {\bibinfo
  {title} {Synthesis, characterization, electronic structure, and phonon
  properties of the noncentrosymmetric superconductor {LaPtSi}},}\ }\href
  {\doibase 10.1103/PhysRevB.88.104508} {\bibfield  {journal} {\bibinfo
  {journal} {Phys. Rev. B}\ }\textbf {\bibinfo {volume} {88}},\ \bibinfo
  {pages} {104508} (\bibinfo {year} {2013})}\BibitemShut {NoStop}%
\bibitem [{\citenamefont {Palazzese}\ \emph {et~al.}(2018)\citenamefont
  {Palazzese}, \citenamefont {Landaeta}, \citenamefont {Subero}, \citenamefont
  {Bauer},\ and\ \citenamefont {Bonalde}}]{palazzese.landaeta.18}%
  \BibitemOpen
  \bibfield  {author} {\bibinfo {author} {\bibfnamefont {S.}~\bibnamefont
  {Palazzese}}, \bibinfo {author} {\bibfnamefont {J.~F.}\ \bibnamefont
  {Landaeta}}, \bibinfo {author} {\bibfnamefont {D.}~\bibnamefont {Subero}},
  \bibinfo {author} {\bibfnamefont {E.}~\bibnamefont {Bauer}}, \ and\ \bibinfo
  {author} {\bibfnamefont {I.}~\bibnamefont {Bonalde}},\ }\bibfield  {title}
  {\enquote {\bibinfo {title} {Strong antisymmetric spin--orbit coupling and
  superconducting properties: the case of noncentrosymmetric {LaPtSi}},}\
  }\href {\doibase 10.1088/1361-648x/aac61d} {\bibfield  {journal} {\bibinfo
  {journal} {J. Phys.: Condens. Matter}\ }\textbf {\bibinfo {volume} {30}},\
  \bibinfo {pages} {255603} (\bibinfo {year} {2018})}\BibitemShut {NoStop}%
\bibitem [{\citenamefont {Domieracki}\ and\ \citenamefont
  {Kaczorowski}(2016)}]{domieracki.kaczorowski.16}%
  \BibitemOpen
  \bibfield  {author} {\bibinfo {author} {\bibfnamefont {K.}~\bibnamefont
  {Domieracki}}\ and\ \bibinfo {author} {\bibfnamefont {D.}~\bibnamefont
  {Kaczorowski}},\ }\bibfield  {title} {\enquote {\bibinfo {title}
  {Superconductivity in a non-centrosymmetric compound {ThCoSi}},}\ }\href
  {\doibase 10.1016/j.jallcom.2016.06.292} {\bibfield  {journal} {\bibinfo
  {journal} {J. Alloy. Comp.}\ }\textbf {\bibinfo {volume} {688}},\ \bibinfo
  {pages} {206} (\bibinfo {year} {2016})}\BibitemShut {NoStop}%
\bibitem [{\citenamefont {Domieracki}\ and\ \citenamefont
  {Kaczorowski}(2018)}]{domieracki.kaczorowski.18}%
  \BibitemOpen
  \bibfield  {author} {\bibinfo {author} {\bibfnamefont {K.}~\bibnamefont
  {Domieracki}}\ and\ \bibinfo {author} {\bibfnamefont {D.}~\bibnamefont
  {Kaczorowski}},\ }\bibfield  {title} {\enquote {\bibinfo {title}
  {Superconductivity in non-centrosymmetric {ThNiSi}},}\ }\href {\doibase
  10.1016/j.jallcom.2017.10.004} {\bibfield  {journal} {\bibinfo  {journal} {J.
  Alloy. Comp.}\ }\textbf {\bibinfo {volume} {731}},\ \bibinfo {pages} {64}
  (\bibinfo {year} {2018})}\BibitemShut {NoStop}%
\bibitem [{\citenamefont {Momma}\ and\ \citenamefont {Izumi}(2011)}]{vesta}%
  \BibitemOpen
  \bibfield  {author} {\bibinfo {author} {\bibfnamefont {K.}~\bibnamefont
  {Momma}}\ and\ \bibinfo {author} {\bibfnamefont {F.}~\bibnamefont {Izumi}},\
  }\bibfield  {title} {\enquote {\bibinfo {title} {{{\sc Vesta3} for
  three-dimensional visualization of crystal, volumetric and morphology
  data}},}\ }\href {\doibase 10.1107/S0021889811038970} {\bibfield  {journal}
  {\bibinfo  {journal} {J. Appl. Crystallogr.}\ }\textbf {\bibinfo {volume}
  {44}},\ \bibinfo {pages} {1272} (\bibinfo {year} {2011})}\BibitemShut
  {NoStop}%
\bibitem [{\citenamefont {Setyawan}\ and\ \citenamefont
  {Curtarolo}(2010)}]{setyawan.curtarolo.10}%
  \BibitemOpen
  \bibfield  {author} {\bibinfo {author} {\bibfnamefont {W.}~\bibnamefont
  {Setyawan}}\ and\ \bibinfo {author} {\bibfnamefont {S.}~\bibnamefont
  {Curtarolo}},\ }\bibfield  {title} {\enquote {\bibinfo {title}
  {High-throughput electronic band structure calculations: Challenges and
  tools},}\ }\href {\doibase 10.1016/j.commatsci.2010.05.010} {\bibfield
  {journal} {\bibinfo  {journal} {Comput. Mater. Sci.}\ }\textbf {\bibinfo
  {volume} {49}},\ \bibinfo {pages} {299} (\bibinfo {year} {2010})}\BibitemShut
  {NoStop}%
\bibitem [{\citenamefont {Bl\"{o}chl}(1994)}]{bloch.94}%
  \BibitemOpen
  \bibfield  {author} {\bibinfo {author} {\bibfnamefont {P.~E.}\ \bibnamefont
  {Bl\"{o}chl}},\ }\bibfield  {title} {\enquote {\bibinfo {title} {Projector
  augmented-wave method},}\ }\href {\doibase 10.1103/PhysRevB.50.17953}
  {\bibfield  {journal} {\bibinfo  {journal} {Phys. Rev. B}\ }\textbf {\bibinfo
  {volume} {50}},\ \bibinfo {pages} {17953} (\bibinfo {year}
  {1994})}\BibitemShut {NoStop}%
\bibitem [{\citenamefont {Perdew}\ \emph {et~al.}(1996)\citenamefont {Perdew},
  \citenamefont {Burke},\ and\ \citenamefont {Ernzerhof}}]{perdew.burke.96}%
  \BibitemOpen
  \bibfield  {author} {\bibinfo {author} {\bibfnamefont {J.~P.}\ \bibnamefont
  {Perdew}}, \bibinfo {author} {\bibfnamefont {K.}~\bibnamefont {Burke}}, \
  and\ \bibinfo {author} {\bibfnamefont {M.}~\bibnamefont {Ernzerhof}},\
  }\bibfield  {title} {\enquote {\bibinfo {title} {Generalized gradient
  approximation made simple},}\ }\href {\doibase 10.1103/PhysRevLett.77.3865}
  {\bibfield  {journal} {\bibinfo  {journal} {Phys. Rev. Lett.}\ }\textbf
  {\bibinfo {volume} {77}},\ \bibinfo {pages} {3865} (\bibinfo {year}
  {1996})}\BibitemShut {NoStop}%
\bibitem [{\citenamefont {Kresse}\ and\ \citenamefont
  {Hafner}(1994)}]{kresse.hafner.94}%
  \BibitemOpen
  \bibfield  {author} {\bibinfo {author} {\bibfnamefont {G.}~\bibnamefont
  {Kresse}}\ and\ \bibinfo {author} {\bibfnamefont {J.}~\bibnamefont
  {Hafner}},\ }\bibfield  {title} {\enquote {\bibinfo {title} {Ab initio
  molecular-dynamics simulation of the liquid-metal--amorphous-semiconductor
  transition in germanium},}\ }\href {\doibase 10.1103/PhysRevB.49.14251}
  {\bibfield  {journal} {\bibinfo  {journal} {Phys. Rev. B}\ }\textbf {\bibinfo
  {volume} {49}},\ \bibinfo {pages} {14251} (\bibinfo {year}
  {1994})}\BibitemShut {NoStop}%
\bibitem [{\citenamefont {Kresse}\ and\ \citenamefont
  {Furthm\"uller}(1996)}]{kresse.furthmuller.96}%
  \BibitemOpen
  \bibfield  {author} {\bibinfo {author} {\bibfnamefont {G.}~\bibnamefont
  {Kresse}}\ and\ \bibinfo {author} {\bibfnamefont {J.}~\bibnamefont
  {Furthm\"uller}},\ }\bibfield  {title} {\enquote {\bibinfo {title} {Efficient
  iterative schemes for ab initio total-energy calculations using a plane-wave
  basis set},}\ }\href {\doibase 10.1103/PhysRevB.54.11169} {\bibfield
  {journal} {\bibinfo  {journal} {Phys. Rev. B}\ }\textbf {\bibinfo {volume}
  {54}},\ \bibinfo {pages} {11169} (\bibinfo {year} {1996})}\BibitemShut
  {NoStop}%
\bibitem [{\citenamefont {Kresse}\ and\ \citenamefont
  {Joubert}(1999)}]{kresse.joubert.99}%
  \BibitemOpen
  \bibfield  {author} {\bibinfo {author} {\bibfnamefont {G.}~\bibnamefont
  {Kresse}}\ and\ \bibinfo {author} {\bibfnamefont {D.}~\bibnamefont
  {Joubert}},\ }\bibfield  {title} {\enquote {\bibinfo {title} {From ultrasoft
  pseudopotentials to the projector augmented-wave method},}\ }\href {\doibase
  10.1103/PhysRevB.59.1758} {\bibfield  {journal} {\bibinfo  {journal} {Phys.
  Rev. B}\ }\textbf {\bibinfo {volume} {59}},\ \bibinfo {pages} {1758}
  (\bibinfo {year} {1999})}\BibitemShut {NoStop}%
\bibitem [{\citenamefont {Parlinski}\ \emph {et~al.}(1997)\citenamefont
  {Parlinski}, \citenamefont {Li},\ and\ \citenamefont
  {Kawazoe}}]{parlinski.li.97}%
  \BibitemOpen
  \bibfield  {author} {\bibinfo {author} {\bibfnamefont {K.}~\bibnamefont
  {Parlinski}}, \bibinfo {author} {\bibfnamefont {Z.~Q.}\ \bibnamefont {Li}}, \
  and\ \bibinfo {author} {\bibfnamefont {Y.}~\bibnamefont {Kawazoe}},\
  }\bibfield  {title} {\enquote {\bibinfo {title} {First-principles
  determination of the soft mode in cubic {ZrO$_{2}$}},}\ }\href {\doibase
  10.1103/PhysRevLett.78.4063} {\bibfield  {journal} {\bibinfo  {journal}
  {Phys. Rev. Lett.}\ }\textbf {\bibinfo {volume} {78}},\ \bibinfo {pages}
  {4063} (\bibinfo {year} {1997})}\BibitemShut {NoStop}%
\bibitem [{\citenamefont {Parlinski}(2018)}]{phonon}%
  \BibitemOpen
  \bibfield  {author} {\bibinfo {author} {\bibfnamefont {K.}~\bibnamefont
  {Parlinski}},\ }\href@noop {} {\enquote {\bibinfo {title} {{\sc Phonon}},}\
  }\bibinfo {address} {Cracow} (\bibinfo {year} {2018})\BibitemShut {NoStop}%
\bibitem [{\citenamefont {Hellman}\ \emph {et~al.}(2013)\citenamefont
  {Hellman}, \citenamefont {Steneteg}, \citenamefont {Abrikosov},\ and\
  \citenamefont {Simak}}]{Hellman2013}%
  \BibitemOpen
  \bibfield  {author} {\bibinfo {author} {\bibfnamefont {O.}~\bibnamefont
  {Hellman}}, \bibinfo {author} {\bibfnamefont {P.}~\bibnamefont {Steneteg}},
  \bibinfo {author} {\bibfnamefont {I.~A.}\ \bibnamefont {Abrikosov}}, \ and\
  \bibinfo {author} {\bibfnamefont {S.~I.}\ \bibnamefont {Simak}},\ }\bibfield
  {title} {\enquote {\bibinfo {title} {Temperature dependent effective
  potential method for accurate free energy calculations of solids},}\ }\href
  {\doibase 10.1103/PhysRevB.87.104111} {\bibfield  {journal} {\bibinfo
  {journal} {Phys. Rev. B}\ }\textbf {\bibinfo {volume} {87}},\ \bibinfo
  {pages} {104111} (\bibinfo {year} {2013})}\BibitemShut {NoStop}%
\bibitem [{\citenamefont {Seo}\ \emph {et~al.}(2012)\citenamefont {Seo},
  \citenamefont {Han},\ and\ \citenamefont {S\'a~de Melo}}]{seo.han.12}%
  \BibitemOpen
  \bibfield  {author} {\bibinfo {author} {\bibfnamefont {K.}~\bibnamefont
  {Seo}}, \bibinfo {author} {\bibfnamefont {L.}~\bibnamefont {Han}}, \ and\
  \bibinfo {author} {\bibfnamefont {C.~A.~R.}\ \bibnamefont {S\'a~de Melo}},\
  }\bibfield  {title} {\enquote {\bibinfo {title} {Topological phase
  transitions in ultracold {Fermi} superfluids: The evolution from
  {Bardeen-Cooper-Schrieffer} to {Bose-Einstein}-condensate superfluids under
  artificial spin-orbit fields},}\ }\href {\doibase 10.1103/PhysRevA.85.033601}
  {\bibfield  {journal} {\bibinfo  {journal} {Phys. Rev. A}\ }\textbf {\bibinfo
  {volume} {85}},\ \bibinfo {pages} {033601} (\bibinfo {year}
  {2012})}\BibitemShut {NoStop}%
\bibitem [{\citenamefont {Ptok}\ \emph {et~al.}(2018)\citenamefont {Ptok},
  \citenamefont {Rodr\'{\i}guez},\ and\ \citenamefont
  {Kapcia}}]{ptok.rodrigez.18}%
  \BibitemOpen
  \bibfield  {author} {\bibinfo {author} {\bibfnamefont {A.}~\bibnamefont
  {Ptok}}, \bibinfo {author} {\bibfnamefont {K.}~\bibnamefont
  {Rodr\'{\i}guez}}, \ and\ \bibinfo {author} {\bibfnamefont {K.~J.}\
  \bibnamefont {Kapcia}},\ }\bibfield  {title} {\enquote {\bibinfo {title}
  {Superconducting monolayer deposited on substrate: Effects of the spin-orbit
  coupling induced by proximity effects},}\ }\href {\doibase
  10.1103/PhysRevMaterials.2.024801} {\bibfield  {journal} {\bibinfo  {journal}
  {Phys. Rev. Materials}\ }\textbf {\bibinfo {volume} {2}},\ \bibinfo {pages}
  {024801} (\bibinfo {year} {2018})}\BibitemShut {NoStop}%
\bibitem [{\citenamefont {Lifshitz}(1960)}]{lifshitz.60}%
  \BibitemOpen
  \bibfield  {author} {\bibinfo {author} {\bibfnamefont {I.~M.}\ \bibnamefont
  {Lifshitz}},\ }\bibfield  {title} {\enquote {\bibinfo {title} {Anomalies of
  electron characteristics of a metal in the high pressure region},}\
  }\href@noop {} {\bibfield  {journal} {\bibinfo  {journal} {Sov. Phys. JETP}\
  }\textbf {\bibinfo {volume} {11}},\ \bibinfo {pages} {1130} (\bibinfo {year}
  {1960})}\BibitemShut {NoStop}%
\end{thebibliography}%

\end{document}